\documentclass[
  aps,
  prx,
  reprint,
  superscriptaddress,
  amsmath,
  amssymb,
  floatfix
]{revtex4-2}

\usepackage{graphicx}
\usepackage{amsmath}
\usepackage{amssymb}
\usepackage{bm}
\usepackage{siunitx}
\usepackage{xcolor}
\usepackage{hyperref}

\begin{document}

\title{Single-electron detection in a high-purity Ge device}

\author{Antoine Armatol}
\author{Corinne Augier}
    \affiliation{IP2I-Lyon, Univ. Lyon, Université Lyon 1, CNRS/IN2P3, Villeurbanne, F-69622, France}
\author{Laurent Berg\'e}
    \affiliation{IJCLab, Université Paris-Saclay, CNRS/IN2P3, Orsay, 91405, France}
\author{Julien Billard}
    \affiliation{IP2I-Lyon, Univ. Lyon, Université Lyon 1, CNRS/IN2P3, Villeurbanne, F-69622, France}
\author{Harvey Birch}
    \affiliation{Department of Physics, University of Zurich, Winterthurerstrasse 190, Zurich, 8057, Switzerland}
\author{Juliette Bl\'e}
    \affiliation{LPSC-IN2P3, Univ. Grenoble Alpes, CNRS, Grenoble INP, Grenoble, 38000, France}
\author{Clarence Chang}
    \affiliation{Argonne National Laboratory, 9700 S Cass Ave, Lemont, IL, 60439, USA}
\author{Yen-Yung Chang}
    \affiliation{Department of Physics, University of California Berkeley, Berkeley, CA, 94720, USA}
\author{Luke Chaplinsky}
    \affiliation{Amherst Center for Fundamental Interactions and Department of Physics, University of Massachusetts, 101 Stockbridge Hall, 80 Campus Center Way, Amherst, MA, 01003-9337, USA}
\author{Gordon Cline}
    \affiliation{Lawrence Berkeley National Laboratory, 1 Cyclotron Rd., Berkeley, CA, 94720, USA}
\author{Alec Cochard}
\author{Ion Cojocari}
    \affiliation{IJCLab, Université Paris-Saclay, CNRS/IN2P3, Orsay, 91405, France}
\author{Jules Colas}
    \affiliation{IP2I-Lyon, Univ. Lyon, Université Lyon 1, CNRS/IN2P3, Villeurbanne, F-69622, France}
\author{Simon Connolly}
   \affiliation{Amherst Center for Fundamental Interactions and Department of Physics, University of Massachusetts, 101 Stockbridge Hall, 80 Campus Center Way, Amherst, MA, 01003-9337, USA}
\author{Maryvonne De Jesus}
    \affiliation{IP2I-Lyon, Univ. Lyon, Université Lyon 1, CNRS/IN2P3, Villeurbanne, F-69622, France}
\author{Pierre de Marcillac}
    \affiliation{IJCLab, Université Paris-Saclay, CNRS/IN2P3, Orsay, 91405, France}
\author{Kamil Dwinger}
    \affiliation{Department of Physics, University of Zurich, Winterthurerstrasse 190, Zurich, 8057, Switzerland}
\author{Romain Faure}
    \affiliation{IP2I-Lyon, Univ. Lyon, Université Lyon 1, CNRS/IN2P3, Villeurbanne, F-69622, France}
\author{Simon Fiorucci}
\author{Maurice Garcia-Sciveres}
    \affiliation{Lawrence Berkeley National Laboratory, 1 Cyclotron Rd., Berkeley, CA, 94720, USA}
\author{Jules Gascon}
    \affiliation{IP2I-Lyon, Univ. Lyon, Université Lyon 1, CNRS/IN2P3, Villeurbanne, F-69622, France}
\author{Ryan Gibbons}
\author{Gil Gilchriese}
    \affiliation{Lawrence Berkeley National Laboratory, 1 Cyclotron Rd., Berkeley, CA, 94720, USA}
\author{Cloé Girard-Carillo}
    \affiliation{LPSC-IN2P3, Univ. Grenoble Alpes, CNRS, Grenoble INP, Grenoble, 38000, France}
\author{Wei Guo}
    \affiliation{FAMU-FSU College of Engineering, Florida State University, 2525 Pottsdamer Street, Tallahassee, FL, 32310, USA}
    \affiliation{National High Magnetic Field Laboratory, 1800 East Paul Dirac Drive, Tallahassee, FL, 32310, USA}
\author{Leila Haegel}
    \affiliation{IP2I-Lyon, Univ. Lyon, Université Lyon 1, CNRS/IN2P3, Villeurbanne, F-69622, France}
\author{Scott Haselschwardt}
    \affiliation{Randall Laboratory of Physics, University of Michigan, Ann Arbor, MI, 48109-1040, USA}
\author{Scott Hertel}
    \affiliation{Amherst Center for Fundamental Interactions and Department of Physics, University of Massachusetts, 101 Stockbridge Hall, 80 Campus Center Way, Amherst, MA, 01003-9337, USA}
\author{Tyler Horoho}
    \affiliation{Randall Laboratory of Physics, University of Michigan, Ann Arbor, MI, 48109-1040, USA}
\author{Keith Hunter}
    \affiliation{Department of Physics and Astronomy, Texas A\&M University, 4242 TAMU, 578 University Dr., College Station, TX, 77843-4242, USA}
\author{Leslie Juigne}
    \affiliation{Department of Physics, University of Zurich, Winterthurerstrasse 190, Zurich, 8057, Switzerland}
\author{Alexandre Juillard}
    \affiliation{IP2I-Lyon, Univ. Lyon, Université Lyon 1, CNRS/IN2P3, Villeurbanne, F-69622, France}
\author{Alex Kavner}
    \affiliation{Department of Physics, University of Zurich, Winterthurerstrasse 190, Zurich, 8057, Switzerland}
\author{Jacob Lamblin}
    \affiliation{LPSC-IN2P3, Univ. Grenoble Alpes, CNRS, Grenoble INP, Grenoble, 38000, France}
\author{Tatiana Le-Bellec}
    \affiliation{IP2I-Lyon, Univ. Lyon, Université Lyon 1, CNRS/IN2P3, Villeurbanne, F-69622, France}
\author{Junsong Lin}
    \affiliation{Department of Physics, University of California Berkeley, Berkeley, CA, 94720, USA}
    \affiliation{Lawrence Berkeley National Laboratory, 1 Cyclotron Rd., Berkeley, CA, 94720, USA}
\author{Rupak Mahapatra}
    \affiliation{Department of Physics and Astronomy, Texas A\&M University, 4242 TAMU, 578 University Dr., College Station, TX, 77843-4242, USA}
\author{Stefanos Marnieros}
    \email{Contact author: stefanos.marnieros@ijclab.in2p3.fr}
    \affiliation{IJCLab, Université Paris-Saclay, CNRS/IN2P3, Orsay, 91405, France}
\author{Claire Marrache-Kikuchi}
    \affiliation{IJCLab, Université Paris-Saclay, CNRS/IN2P3, Orsay, 91405, France}
\author{Nicolas Martini}
    \affiliation{IP2I-Lyon, Univ. Lyon, Université Lyon 1, CNRS/IN2P3, Villeurbanne, F-69622, France}
\author{Will Matava}
\author{Daniel McKinsey}
    \affiliation{Lawrence Berkeley National Laboratory, 1 Cyclotron Rd., Berkeley, CA, 94720, USA}
\author{Connor McLaughlin}
    \affiliation{Amherst Center for Fundamental Interactions and Department of Physics, University of Massachusetts, 101 Stockbridge Hall, 80 Campus Center Way, Amherst, MA, 01003-9337, USA}
\author{Johann Menu}
    \affiliation{LPSC-IN2P3, Univ. Grenoble Alpes, CNRS, Grenoble INP, Grenoble, 38000, France}
\author{Knut Moraa}
    \affiliation{Department of Physics, University of Zurich, Winterthurerstrasse 190, Zurich, 8057, Switzerland}
\author{Valentina Novati}
    \affiliation{LPSC-IN2P3, Univ. Grenoble Alpes, CNRS, Grenoble INP, Grenoble, 38000, France}
\author{Emiliano Olivieri}
    \affiliation{IJCLab, Université Paris-Saclay, CNRS/IN2P3, Orsay, 91405, France}
\author{Björn Penning}
    \affiliation{Department of Physics, University of Zurich, Winterthurerstrasse 190, Zurich, 8057, Switzerland}
\author{Mark Platt}
    \affiliation{Department of Physics and Astronomy, Texas A\&M University, 4242 TAMU, 578 University Dr., College Station, TX, 77843-4242, USA}
\author{Denys Poda}
    \affiliation{IJCLab, Université Paris-Saclay, CNRS/IN2P3, Orsay, 91405, France}
\author{Matt Pyle}
    \affiliation{Department of Physics, University of California Berkeley, Berkeley, CA, 94720, USA}
\author{Yinghe Qi}
    \affiliation{FAMU-FSU College of Engineering, Florida State University, 2525 Pottsdamer Street, Tallahassee, FL, 32310, USA}
    \affiliation{National High Magnetic Field Laboratory, 1800 East Paul Dirac Drive, Tallahassee, FL, 32310, USA}
\author{Rafsan Rabbi}
    \affiliation{FAMU-FSU College of Engineering, Florida State University, 2525 Pottsdamer Street, Tallahassee, FL, 32310, USA}
\author{Ivar Rydstrom}
\author{Bernard Sadoulet}
    \affiliation{Department of Physics, University of California Berkeley, Berkeley, CA, 94720, USA}
\author{Silvia Scorza}
    \affiliation{LPSC-IN2P3, Univ. Grenoble Alpes, CNRS, Grenoble INP, Grenoble, 38000, France}
\author{Bruno Serfass}
    \affiliation{Department of Physics, University of California Berkeley, Berkeley, CA, 94720, USA}
\author{Peter Sorensen}
    \affiliation{Lawrence Berkeley National Laboratory, 1 Cyclotron Rd., Berkeley, CA, 94720, USA}
\author{Gabrielle Soum-Sidikov}
    \affiliation{IP2I-Lyon, Univ. Lyon, Université Lyon 1, CNRS/IN2P3, Villeurbanne, F-69622, France}
\author{Samara Steinfeld}
    \affiliation{Lawrence Berkeley National Laboratory, 1 Cyclotron Rd., Berkeley, CA, 94720, USA}
\author{Henry Su}
    \affiliation{Amherst Center for Fundamental Interactions and Department of Physics, University of Massachusetts, 101 Stockbridge Hall, 80 Campus Center Way, Amherst, MA, 01003-9337, USA}
\author{Toki Suzuki}
    \affiliation{Lawrence Berkeley National Laboratory, 1 Cyclotron Rd., Berkeley, CA, 94720, USA}
\author{Ronald Vaughn II}
    \affiliation{Amherst Center for Fundamental Interactions and Department of Physics, University of Massachusetts, 101 Stockbridge Hall, 80 Campus Center Way, Amherst, MA, 01003-9337, USA}
\author{Charlie Veihmeyer}
    \affiliation{Department of Physics, University of California Berkeley, Berkeley, CA, 94720, USA}
\author{Paul Vittaz}
    \affiliation{IP2I-Lyon, Univ. Lyon, Université Lyon 1, CNRS/IN2P3, Villeurbanne, F-69622, France}
\author{Gensheng Wang}
    \affiliation{Argonne National Laboratory, 9700 S Cass Ave, Lemont, IL, 60439, USA}
\author{Yue Wang}
\author{Michael R. Williams}
    \affiliation{Lawrence Berkeley National Laboratory, 1 Cyclotron Rd., Berkeley, CA, 94720, USA}
\author{Joanna Wuko}
    \affiliation{Amherst Center for Fundamental Interactions and Department of Physics, University of Massachusetts, 101 Stockbridge Hall, 80 Campus Center Way, Amherst, MA, 01003-9337, USA}
\author{Volodymyr Yefremenko}
    \affiliation{Argonne National Laboratory, 9700 S Cass Ave, Lemont, IL, 60439, USA}

\collaboration{TESSERACT Collaboration}

\author{Alexandre Broniatowski}
\author{Louis Dumoulin}
\affiliation{IJCLab, Université Paris-Saclay, CNRS/IN2P3, Orsay, 91405, France}

\date{\today}








    





\begin{abstract}
High-purity germanium (HPGe) detectors are among the most powerful instruments for gamma ray spectroscopy, combining exceptional energy resolution with large active volumes. Although they operate close to the fundamental resolution limit at high energies, their performance at very low energies has long been limited by electronic noise, restricting detection thresholds to above approximately 30 electron-hole pairs ($\sim$100~eV). Here we introduce a cryogenic HPGe detector architecture that achieves single electron-hole pair sensitivity by calorimetrically measuring ionization at temperatures near 20 mK. The device integrates a NbSi transition-edge sensor into a point-contact-inspired geometry, concentrating athermal phonon energy generated during charge drift and enabling eV-scale sensitivity ultimately set by the semiconductor band gap. Beyond single-charge detection, the device rejects the low-energy excess background that limits existing cryogenic low-threshold technologies, establishing a pathway towards discrimination between electron- and nuclear-recoil events below 100~eV. These advances open a route towards next-generation detectors for direct dark matter searches, coherent elastic neutrino-nucleus scattering, and nuclear reactor monitoring for non-proliferation applications.
\end{abstract}




\maketitle

\section{Introduction}

High-purity germanium (HPGe) devices build on a well-established and highly mature technology for gamma-ray detection, which has proven particularly effective in experiments demanding large active detector volumes \cite{Ichimura:2023,Burlac:2025,Geng:2024,Armengaud:2017,Aralis:2020,Aralis:2021,Adamski:2025,Ackermann:2025,Armatol:2025}. 
Thanks to decades of progress in the growth and purification of germanium crystals, the concentration of residual impurities can now be reduced below $10^{10}$~cm$^{-3}$, allowing the stable operation of kilogram-scale detectors. State-of-the-art HPGe devices are approaching the resolution limit set by the Fano factor \cite{Fano:1947,Bilger:1967,Lowe:1997}, on the order of 500~eV at 100~keV, and achieve energy thresholds close to 100 eV \cite{Ackermann:2024,Augier:2024}, the latter being mainly restricted by the readout noise from the charge amplifier. In germanium, the average energy required to produce a single electron-hole (e-h) pair is approximately 3~eV for gamma quanta \cite{Antman:1966,Wei:2017}, corresponding to thresholds of about 30~e-h pairs.

Recent experimental efforts aimed at the direct detection of light dark matter and the observation of coherent elastic neutrino-nucleus scattering (CE$\nu$NS) have created a strong motivation to further improve the resolution of large-mass semiconductor detectors, ultimately targeting the detection of single e-h pair events \cite{Agnese:2018,Arnaud:2020,Mei:2024}. Since the expected dark matter and CE$\nu$NS recoil spectra rise steeply at low energies, reaching eV-scale thresholds would provide unprecedented sensitivity and open a new window to explore physics beyond the Standard Model.

For silicon-based semiconductors, two technologies employing crystals with masses of 1--10 g have demonstrated single e-h pair sensitivity. The first approach is based on a charge-coupled device (CCD) architecture, where the incident radiation is integrated over a long exposure prior to a sequential readout of each pixel charge using an innovative technique. These devices, known as Skipper-CCDs \cite{Tiffenberg:2017}, combine single e-h pair thresholds, sub-0.1 charge rms resolution, and extremely low dark-current backgrounds \cite{Bloch:2025,Aggarwal:2025}. By operating dozens of Skipper-CCDs in a tower configuration, a total target mass on the order of a kilogram can be achieved.
A second technology capable of single-charge sensitivity exploits an alternative ionization-readout mechanism based on the Neganov-Trofimov-Luke (NTL) effect \cite{Neganov:1985,Luke:1988}. In this approach, the ionization signal is inferred from a low-temperature phonon sensor that measures the athermal phonons generated as e-h pairs drift through the crystal. The energy released during the drift of $N$ e-h pairs scales linearly with the voltage $V_\textrm{NTL}$ applied across the semiconductor crystal, yielding $E_\textrm{NTL}=N\cdot V_\textrm{NTL}$, where $E_\textrm{NTL}$ is expressed in electronvolts. For biases of order 35--100~V, the NTL contribution exceeds the initial particle-deposition by more than an order of magnitude, enabling baseline resolutions below 0.1~e-h pair (rms) in gram-scale silicon detectors instrumented with transition-edge sensors (TES) \cite{Agnese:2018,Kennard:2026}. Moreover, these devices exhibit substantial suppression of low-energy excess events \cite{Albakry:2025,Albakry:2026}, which often dominate the background in low-threshold cryogenic detectors \cite{Adari:2022,Baxter:2025,Anthony-Petersen:2025}.

Single e-h pair detection is also attainable in smaller semiconductor devices through charge-multiplication processes based on impact ionization and avalanche effects. Single-photon avalanche photodiodes (APDs) operating in the so-called Geiger mode can intrinsically amplify single e-h excitations into measurable pulse signals \cite{Cusini:2022,Liu:2024}. High-efficiency Ge-target devices can be realized by epitaxially growing a thin Ge layer on a Si-APD substrate (Ge-on-Si APD) \cite{Na:2024}. Although charge-multiplication techniques compatible with large-mass Ge crystals are currently being investigated \cite{Acharya:2023,Mei:2024}, they have not yet reached the expected performance levels.

Here we present a low-temperature HPGe detector that combines NTL-assisted single-charge detection with a novel point-contact TES architecture. Scaling NTL-based detectors to large absorber masses presents a fundamental challenge: in conventional geometries, NTL-athermal phonons generated throughout the crystal must propagate to sensors covering a substantial fraction of its surface, making efficient phonon collection increasingly difficult as the detector dimensions increase. Our approach circumvents this limitation by shaping the electric field such that most of the NTL phonon energy is generated locally, in the immediate vicinity of a small and highly sensitive TES. With this architecture, we demonstrate single e-h pair sensitivity in a 40-g Ge crystal and establish a pathway toward single-charge detection in substantially larger, potentially kilogram-scale, HPGe detectors. An independent thermal-phonon sensor provides complementary calorimetric information, enabling rejection of a substantial fraction of the low-energy background and opening a route toward event-by-event particle discrimination.

\section{Detector concept and operating principle}

\begin{figure*}[!ht]
  \centering
  \includegraphics[height=0.31\textheight]{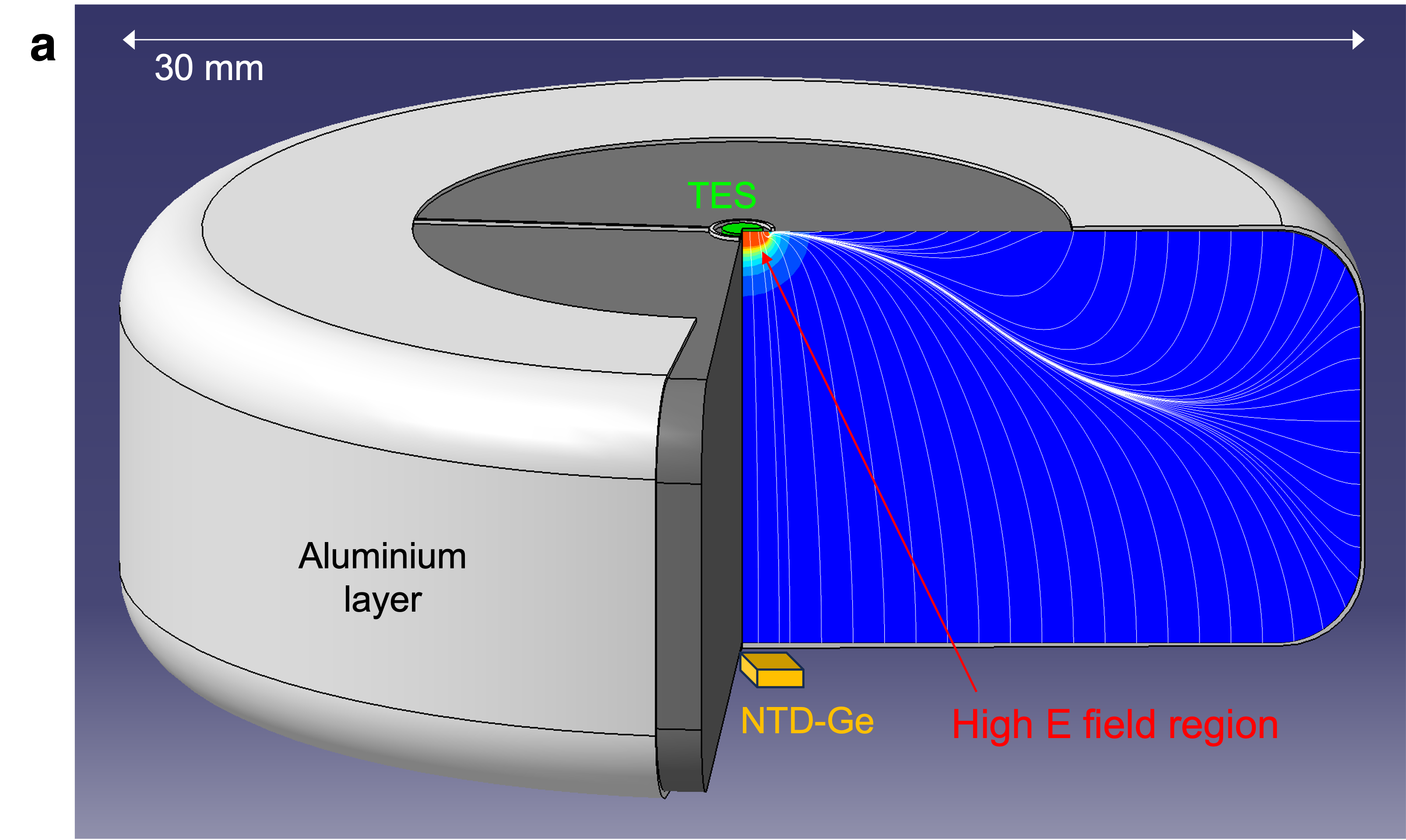}
  \includegraphics[height=0.24\textheight]{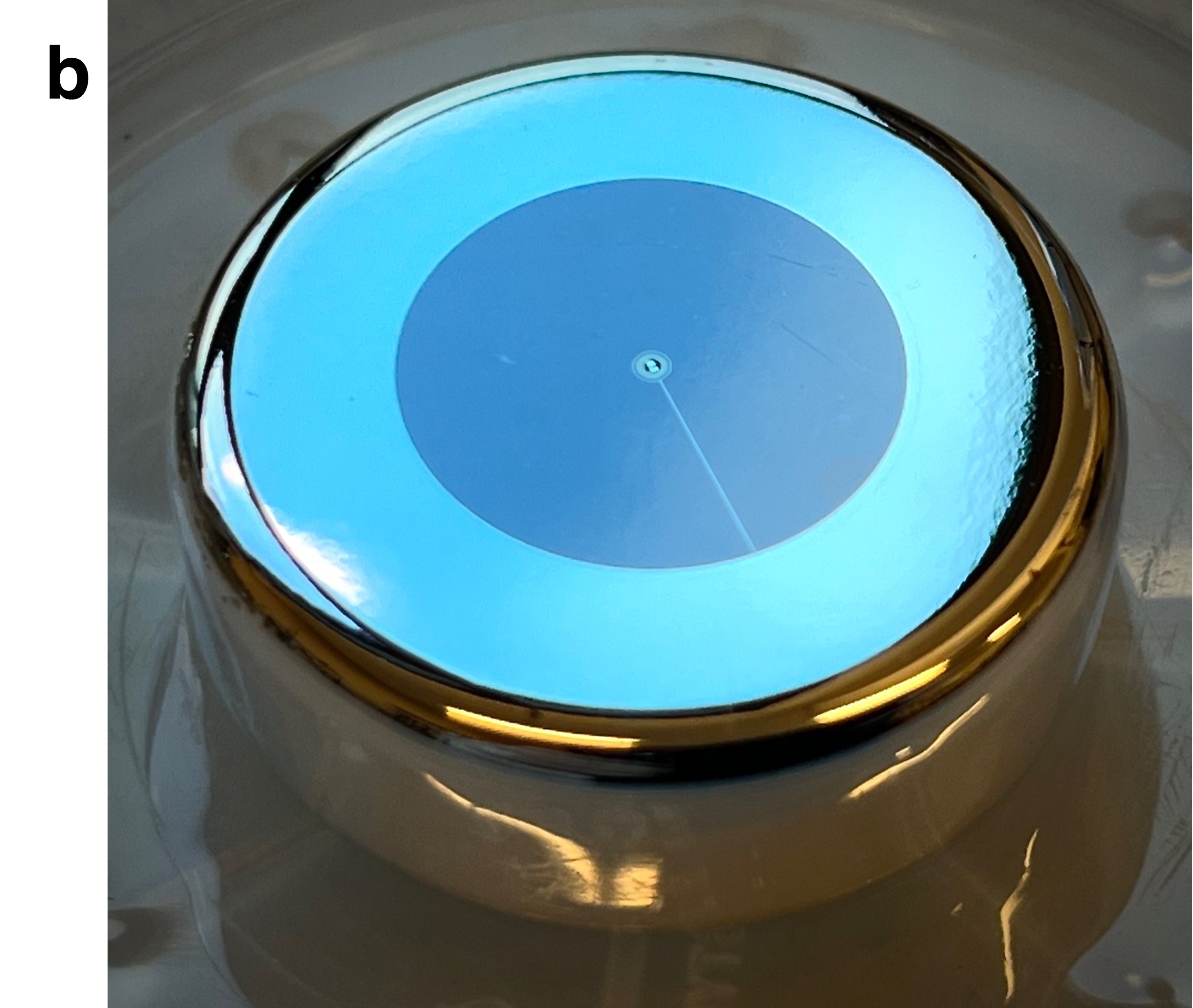}
  \includegraphics[height=0.245\textheight]{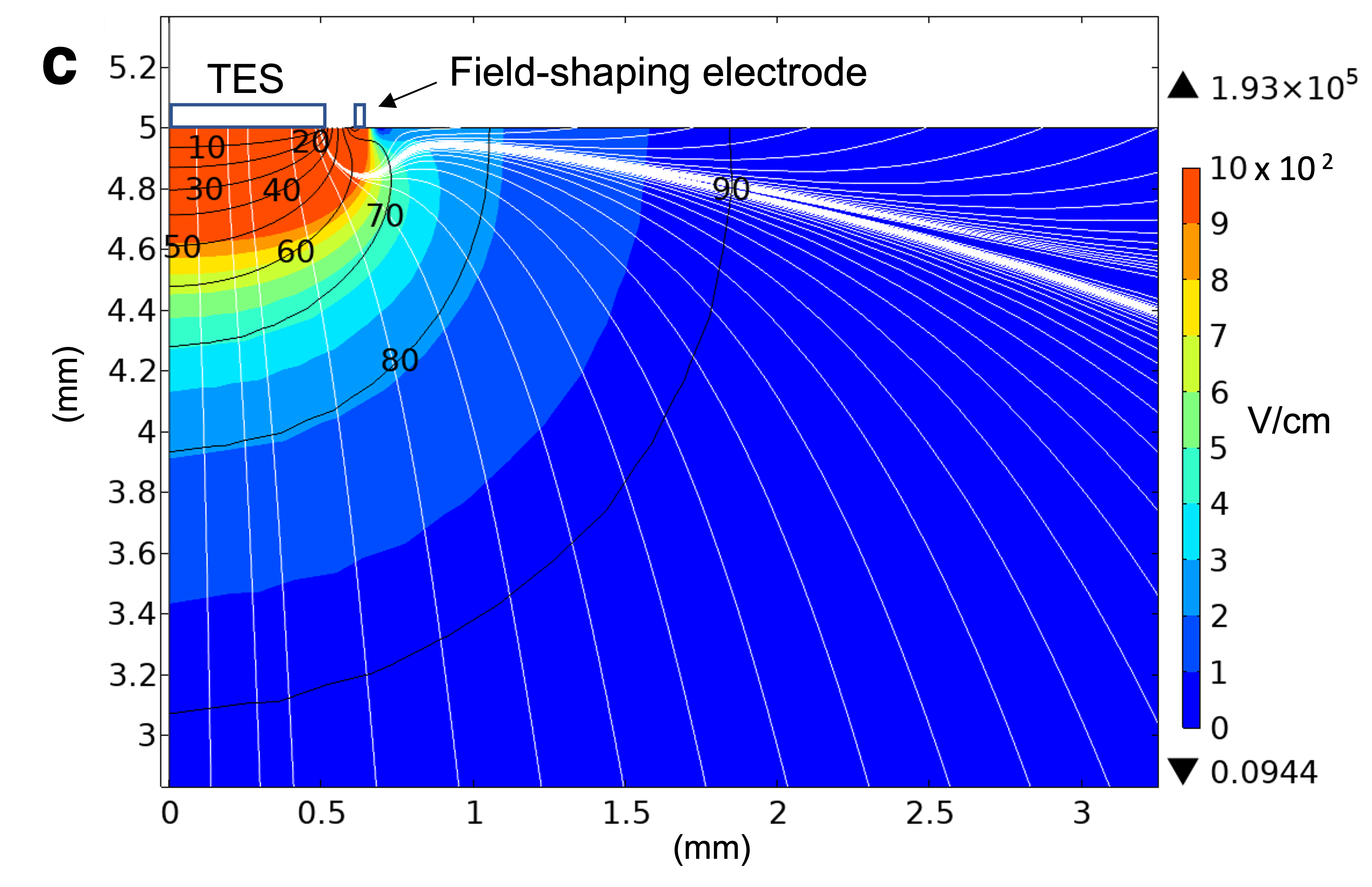}
     \caption{Detector design.
     (a)~Point-contact TES detector with simulated electric-field lines (white) showing a strong field region (red) beneath the TES. The Al electrode covers the bottom, lateral, and part of the top crystal surfaces and is connected to a field-shaping ring surrounding the TES (green) by a 16-$\mu$m-wide line.
     (b)~Photograph of the detector top surface.
     (c)~TES-region simulation at $V_\textrm{NTL}=$~100~V. The red region corresponds to electric fields exceeding 900 V/cm, with peak fields above $10^{5}$~V/cm. Simulated equipotential contours (black) show that more than 80\% of the potential drop occurs within 1~mm of the TES, implying that a corresponding fraction of the NTL athermal phonons is generated in this region. Electric-field lines (white) show a clear deviation near the TES due to the Al field-shaping electrode.}
  \label{fig:Detector}
\end{figure*}

The detector consists of a cylindrical Ge crystal instrumented with two thermal sensors, a NbSi TES and a neutron-transmutation-doped germanium (NTD-Ge) thermistor, placed on opposite faces. Charge collection is achieved by applying an electric field across the Ge crystal via an aluminium electrode covering most of the surface (Fig.~\ref{fig:Detector}). The TES is optimized to measure the ionization signal with high precision by detecting athermal phonons generated through the NTL effect. To enhance sensitivity, the device employs an innovative point-contact geometry: in contrast to conventional detectors, the point-contact electrode on the top surface is replaced by the TES itself. When a bias voltage $V_\textrm{NTL}$ is applied to the Al electrode, a region of strong electric field forms in the Ge near the TES, which is held at ground potential. Figures~\ref{fig:Detector}(a) and \ref{fig:Detector}(c) illustrate the simulated electric field lines and equipotential contours, obtained with the COMSOL Multiphysics finite-element software \cite{COMSOL}. Charge carriers produced by a particle interaction drift across the crystal and reach the TES or the Al electrode within less than 1~$\mu$s. During this drift, athermal phonons are emitted predominantly in the high-field region, where they couple efficiently to the TES and substantially enhance its response, largely independent of the interaction position. Owing to the point-contact geometry, most of the athermal phonon energy is deposited within approximately 1~mm of the TES, irrespective of the absorber dimensions. This localization enables scaling of the Ge absorber to larger volumes while preserving the TES signal amplitude.
For particle interactions occurring away from the immediate vicinity of the TES, athermal phonons generated by the recoiling electron or nucleus are predominantly absorbed by the superconducting Al layer covering most of the crystal surface. Consequently, their contribution to the TES signal is negligible compared with that of the NTL phonons. Although the point-contact geometry does not enhance the response of the NTD-Ge calorimeter, which primarily measures the absorber's global temperature rise, the combined operation of the TES and NTD-Ge sensors provides robust discrimination against non-ionizing events, a dominant source of low-energy background in cryogenic rare-event searches. In addition, the NTD-Ge sensor enables reliable energy calibration above 1~keV while continuously monitoring the charge-collection efficiency. Details of the detector fabrication and calibration procedures are provided in the Methods.

\section{Single-electron sensitivity}

The TES sensitivity was calibrated using a 1590-nm (0.78~eV) pulsed laser coupled to the detector through a single-mode optical fiber. The laser delivered 10-ns pulses at a repetition rate of 10~Hz, producing signals spanning the 0–30~eV energy range. At this wavelength, the photon absorption length in Ge is approximately 4.5~mm \cite{Macfarlane:1974}, ensuring that photons entering the crystal interact predominantly within the bulk. Although an average energy of approximately 3~eV is required to create an \mbox{e-h} pair in Ge at high energies, the 0.78~eV photons lie above the Ge band gap and can therefore generate e-h excitations. Photon transmission into the Ge absorber was optimized by aligning the optical fiber with the top surface not covered by the Al electrode. Before cool-down, the detector was irradiated with a $^{252}$Cf neutron source, producing $^{71}$Ge isotopes uniformly throughout the crystal volume. The subsequent decay of $^{71}$Ge to $^{71}$Ga yields K-, L-, and M-shell X-rays at 10.37~keV, 1.3~keV, and 160~eV, respectively \cite{Genz:1971,Norman:2024}, providing well-defined internal calibration lines for the detector.
Before each data-taking period, a regeneration procedure was performed to neutralize bulk and surface impurities and suppress residual space charge (see Appendix~\ref{App_SpaceCharge}). Data were acquired with NTL bias voltages ranging from 0 to 90 V and a TES bias voltage of 0.6~$\mu$V (see Appendix~\ref{App_Calibration}). NTL biases of 50--70~V provided the optimal compromise between signal-to-noise ratio and detector leakage current.

\begin{figure*}[!ht]
  \centering
  \includegraphics[height=0.31\textheight]{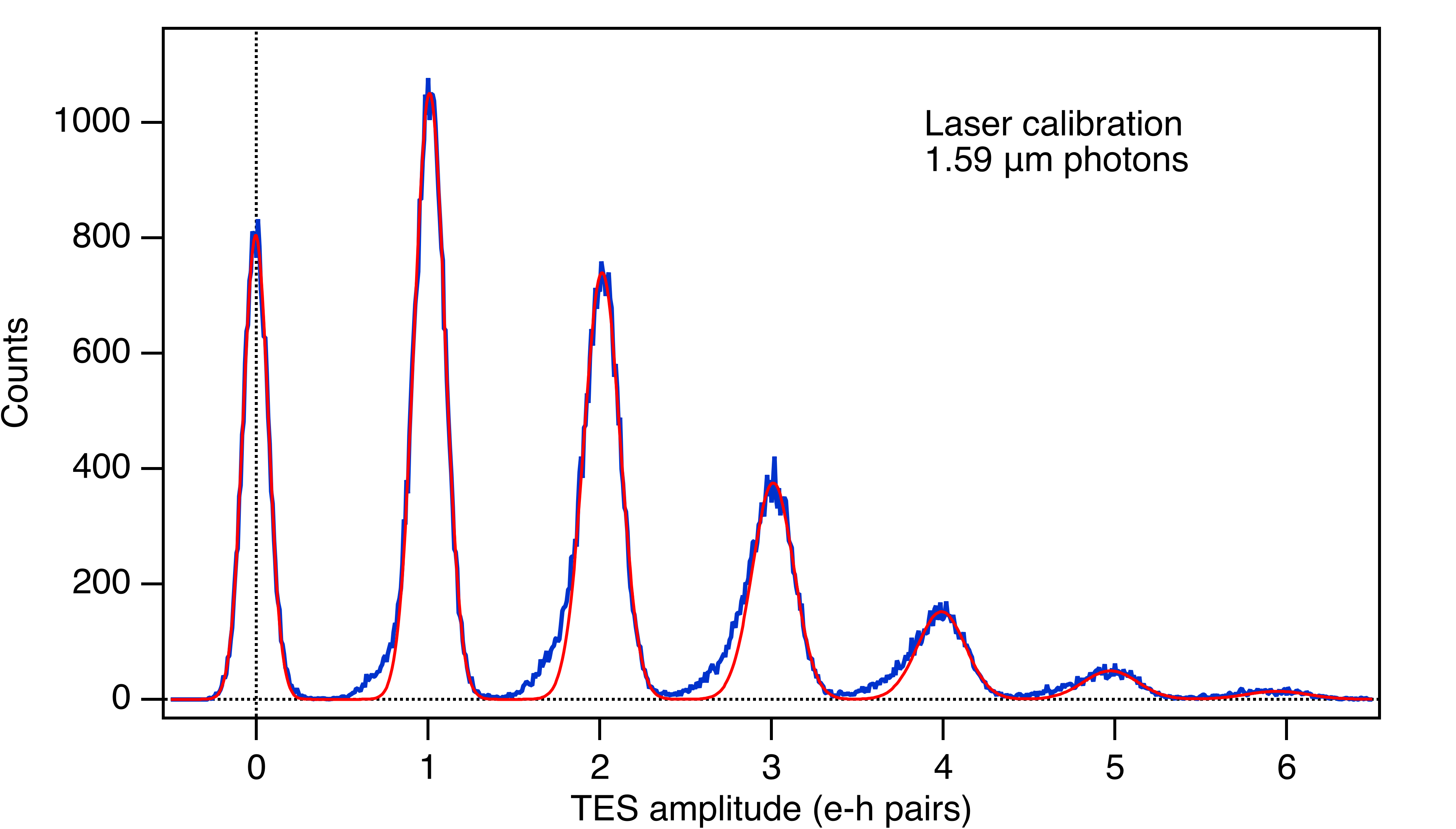}
    \caption{Single electron-hole resolution. Spectrum of TES signal amplitudes (blue), for laser-induced events at an NTL voltage of 50~V. The red curve shows Gaussian fits to the peaks.}
  \label{fig:TES_eh_peaks}
\end{figure*}

Figure \ref{fig:TES_eh_peaks} shows the distribution of fitted TES pulse amplitudes obtained during laser calibration at an NTL voltage of 50~V. The spectrum exhibits a series of well-resolved peaks corresponding to discrete numbers of e-h pairs generated in the Ge crystal. This multi-peak structure arises from Poisson fluctuations in the number of photons emitted per laser pulse, together with the probability that an absorbed photon generates an e-h excitation in Ge. To first order, the data are well described by a Poisson distribution with a mean of 1.7 detected photons per pulse convolved with a Gaussian broadening arising from the finite phonon-energy resolution of the TES. To our knowledge, this represents the first demonstration of an HPGe detector achieving single e-h pair sensitivity with a baseline resolution below 0.1~e-h pairs (rms).

The baseline noise, set by combined SQUID readout and intrinsic TES noise, is independent of the NTL bias. No excess noise is observed as the NTL voltage increases from 0 to 90 V, leading to a monotonic improvement in signal-to-noise ratio and a baseline resolution of 0.06 e-h pairs at 90 V. Above $\sim$70 V, an additional background of single e-h events appears, attributed to leakage-current-driven charge injection. For applied voltages exceeding 120~V, the leakage current can trigger a self-sustained avalanche process in the high-field region of the crystal, heating the detector above 200~mK. Under negative bias applied to the Al electrode, avalanche onset occurs at substantially lower voltages, of order $-$20~V. All data presented here were acquired with positive bias on the Al electrode, corresponding to hole collection at the TES.

At higher energies, additional contributions to the e-h peak broadening become evident, which are consistent with charge trapping during carrier drift and fluctuations in the generation and propagation of NTL-induced athermal phonons. The cumulative effect  of these processes ultimately limits single-electron sensitivity, with individual e-h peaks no longer resolvable above approximately 20~e-h pairs in the present device. 
 
Below this limit, the e-h peaks develop a low-energy tail that deviates from a purely Gaussian profile (Fig.~\ref{fig:TES_eh_peaks}). Although the Gaussian baseline resolution continues to improve with increasing NTL bias owing to enhanced athermal-phonon emission, the fraction of events populating the low-energy tail remains essentially unchanged. Low- and high-energy tails on single-electron peaks have previously been reported in Si phonon-sensing detectors and successfully explained by charge trapping and impact ionization \cite{Ponce:2020,Wilson:2024}. In our detector, however, the specific point-contact design suggests a different origin. We attribute the observed peak asymmetry primarily to the reduced sensitivity of the TES to events in which charge is collected on the bonding pads rather than directly on the TES.
These events occur predominantly for interactions near the crystal symmetry axis, where charge carriers drift toward the TES bonding pads (see Appendix~\ref{App_Fabrication}).

For bias voltages $V_\textrm{NTL}\gtrsim 50$~V, the Gaussian component of the e-h peaks exhibits an rms width that depends only on the phonon energy transmitted to the TES, $E_\textrm{TES}$, and is well described by 
\begin{equation}
    \Delta E_\textrm{TES}=0.325 + 0.016  \cdot E_\textrm{TES},
    \label{Eq_DE_TES}
\end{equation}
where both $E_\textrm{TES}$ and $\Delta E_\textrm{TES}$ are expressed in eV. The quantity $E_\textrm{TES}$ is determined using the electrothermal-feedback self-calibration procedure described in Appendix~\ref{App_Calibration}.

At $V_\textrm{NTL}=50~V$ (Fig.~\ref{fig:TES_eh_peaks}), the corresponding resolution $\Delta Q$ of individual e-h peaks with charge $Q$, expressed in units of e-h pairs, is given by
\begin{equation}
    \Delta Q =0.073 + 0.016  \cdot Q.
\end{equation}
The constant term reflects the TES baseline resolution, while the linear term indicates a progressive degradation with increasing collected charge. This behavior is attributed to fluctuations in athermal-phonon down-conversion and propagation through the crystal, which modulate the fraction of NTL energy collected by the TES. At lower bias voltages ($V_\textrm{NTL}\lesssim 30$~V), residual space charge in the Ge crystal further degrades the energy resolution described by equation \eqref{Eq_DE_TES}.

Figure \ref{fig:Background_plots}(a) shows the TES spectrum over an extended range for acquisition runs at a collection voltage of 70~V. The dynamic range allows simultaneous resolution of single e-h peaks from background events and the K-, L- and M-shell X-ray lines from $^{71}$Ge decay. The 160~eV M-shell line has an rms energy resolution of ($8.5\pm1.7$)~eV, consistent with statistical fluctuations in e-h pair creation corresponding to a Fano factor of $0.15\pm0.07$, in agreement with previous measurements below 1~keV \cite{Arnaud:2020,Lowe:1997}. No discrete e-h peak structure is observed at this energy, indicating that individual charges are no longer resolved. The rms resolutions of the 1.30~keV and 10.37~keV lines are approximately 120~eV and 1.1~keV, respectively, substantially exceeding the Fano limit. Beyond charge trapping and variations in athermal phonon collection efficiency, this degradation arises from reduced TES sensitivity as high energy events drive the sensor above its superconducting transition temperature.

\section{Low-energy background}
                     
\begin{figure*}[!ht]
  \centering
  \includegraphics[height=0.31\textheight]{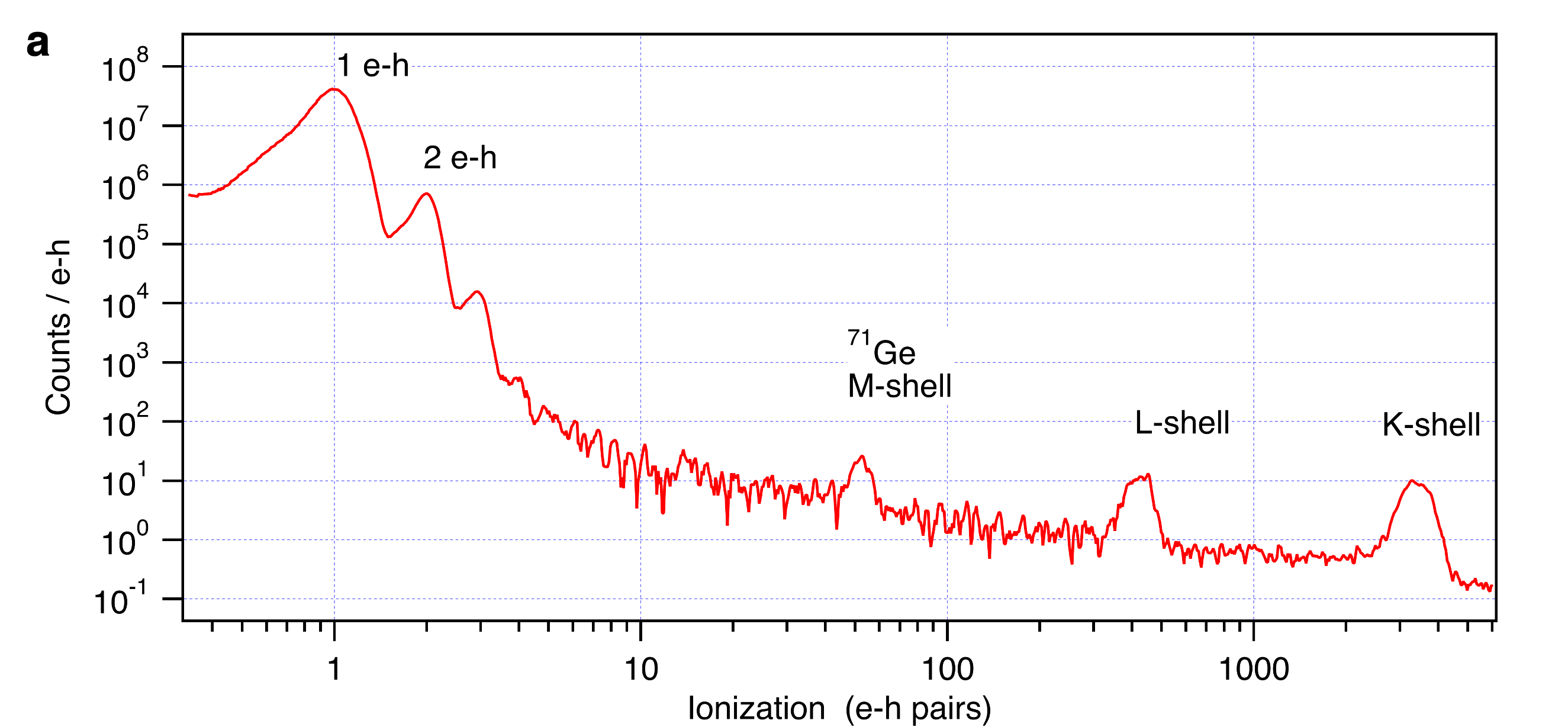}
  \includegraphics[height=0.2\textheight]{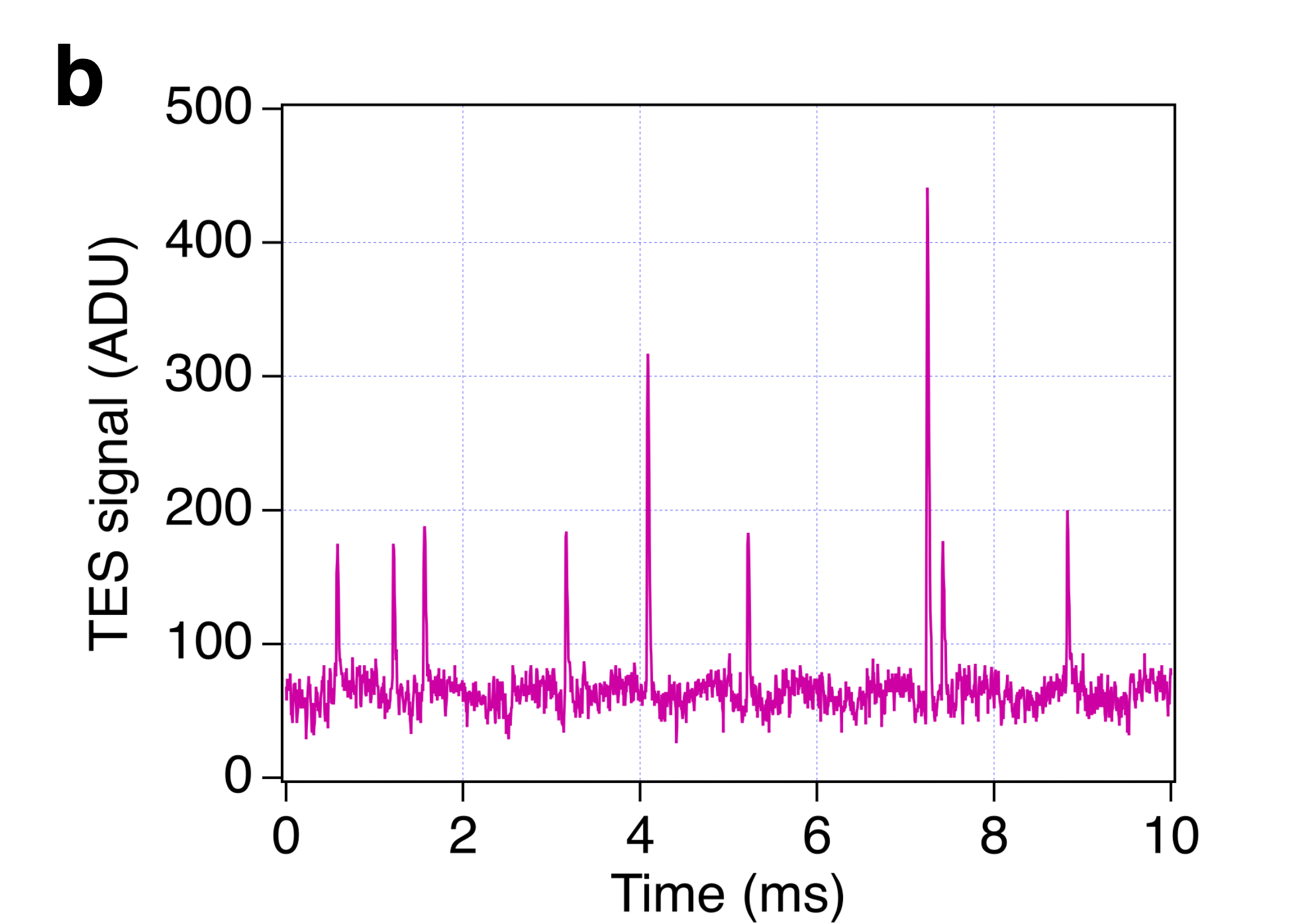}
  \includegraphics[height=0.195\textheight]{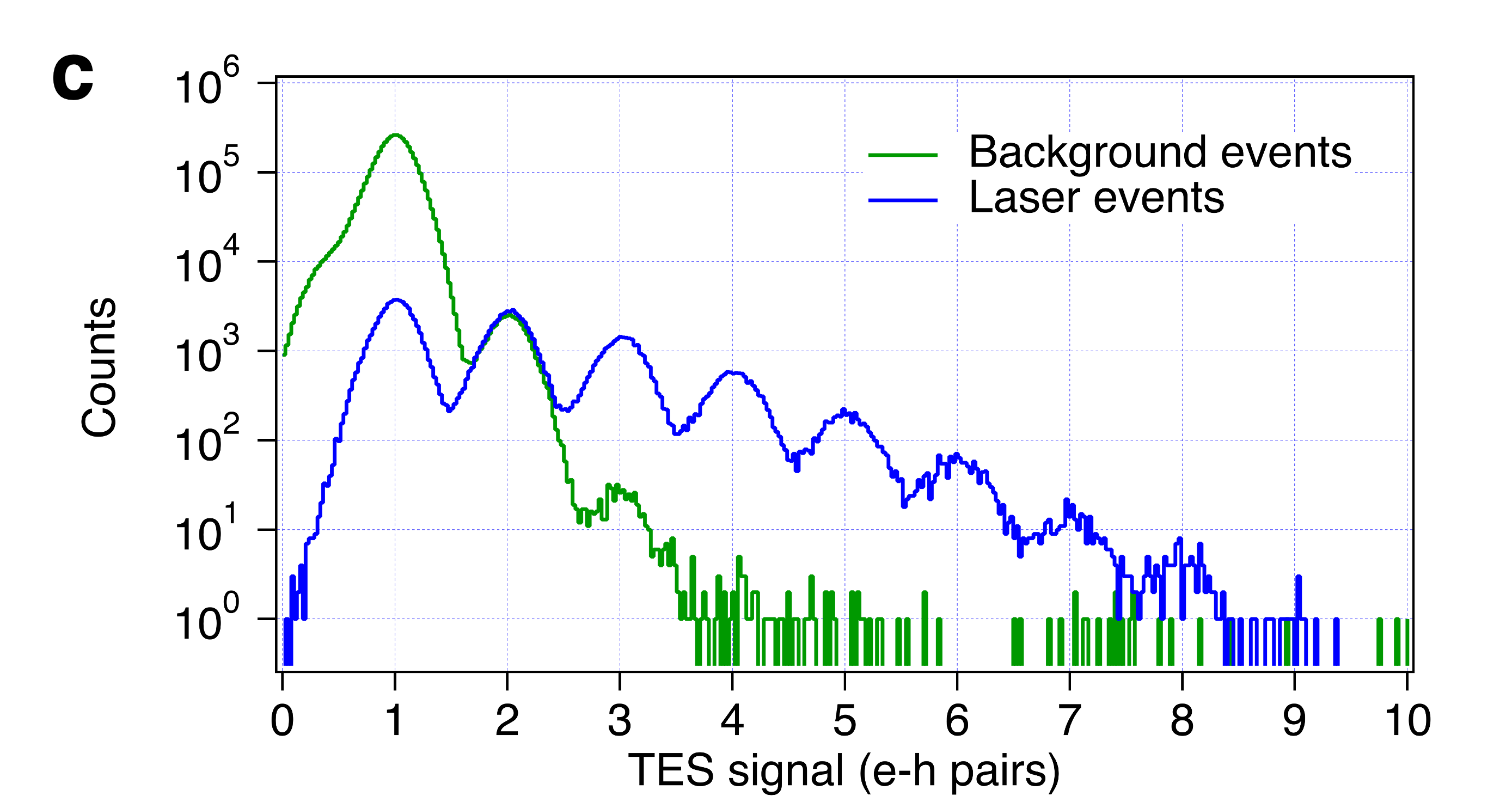}
     \caption{(a) TES energy spectrum of the detector background, including $^{71}$Ge X-ray decays, obtained from 8 h 20 min of combined data from six runs at a 70 V. To accommodate resolution degradation at high energies, the spectrum is plotted with exponentially increasing bin widths (0.003--30~e-h); counts are shown per e-h interval.
     (b) Raw TES data stream from a background acquisition run at a Ge bias of 50~V. Several single e-h pair pulses are visible, along with signals corresponding to 2~e-h and 3~e-h pair events.
     (c) TES-measured spectra from a 2 h calibration run at 50 V, showing background events (green curve) and laser events (blue curve). TES signals are extracted from pulse integrals as described in Appendix~\ref{App_Calibration}.}
  \label{fig:Background_plots}
\end{figure*}

The background spectrum of the device was characterized in detail for NTL voltages between 0 and 90~V, with particular emphasis on the low-energy regime below 10~e-h excitations. A representative 10 ms segment of raw TES data is shown in Fig.~\ref{fig:Background_plots}(b), illustrating a high rate of low-energy background events dominated by single e-h pulses. This behavior is further reflected in Fig.~\ref{fig:Background_plots}(c), which shows the TES energy spectrum accumulated over a 2~h acquisition. A prominent single e-h peak is observed, superimposed on laser-induced events identified via the laser-electronics TTL trigger. The second and third e-h background peaks are less populated but remain clearly resolved. This low-energy background, on the order of 500~Hz, shows no dependence on the applied Ge bias for $V_\textrm{NTL}\lesssim 70$~V and is primarily attributed to far-infrared photons producing single-charge excitations in the Ge absorber, as also suggested by studies on Si cryogenic detectors \cite{Agnese:2018,Romani:2018}. 
In the present setup, the most plausible source of these photons is black-body infrared emission from within the cryostat, which can ionize residual bulk or surface impurities in the Ge crystal.

Additional detector-intrinsic background contributions become apparent at NTL voltages $\gtrsim$~70~V. These include electric-field-assisted tunnel ionization of Ge impurities \cite{Zurauskas:1992}, as well as charge  injection associated with leakage currents from the TES or Al electrode layers.

The observed rate of low-energy background events represents an important limitation for light dark-matter direct detection and CE$\nu$NS experiments, motivating the development of improved suppression of stray infrared radiation within the cryostat, together with background discrimination at the level of single e-h excitations. 

\section{Background discrimination}

\begin{figure*}[!ht]
  \centering
  \includegraphics[width=\linewidth]{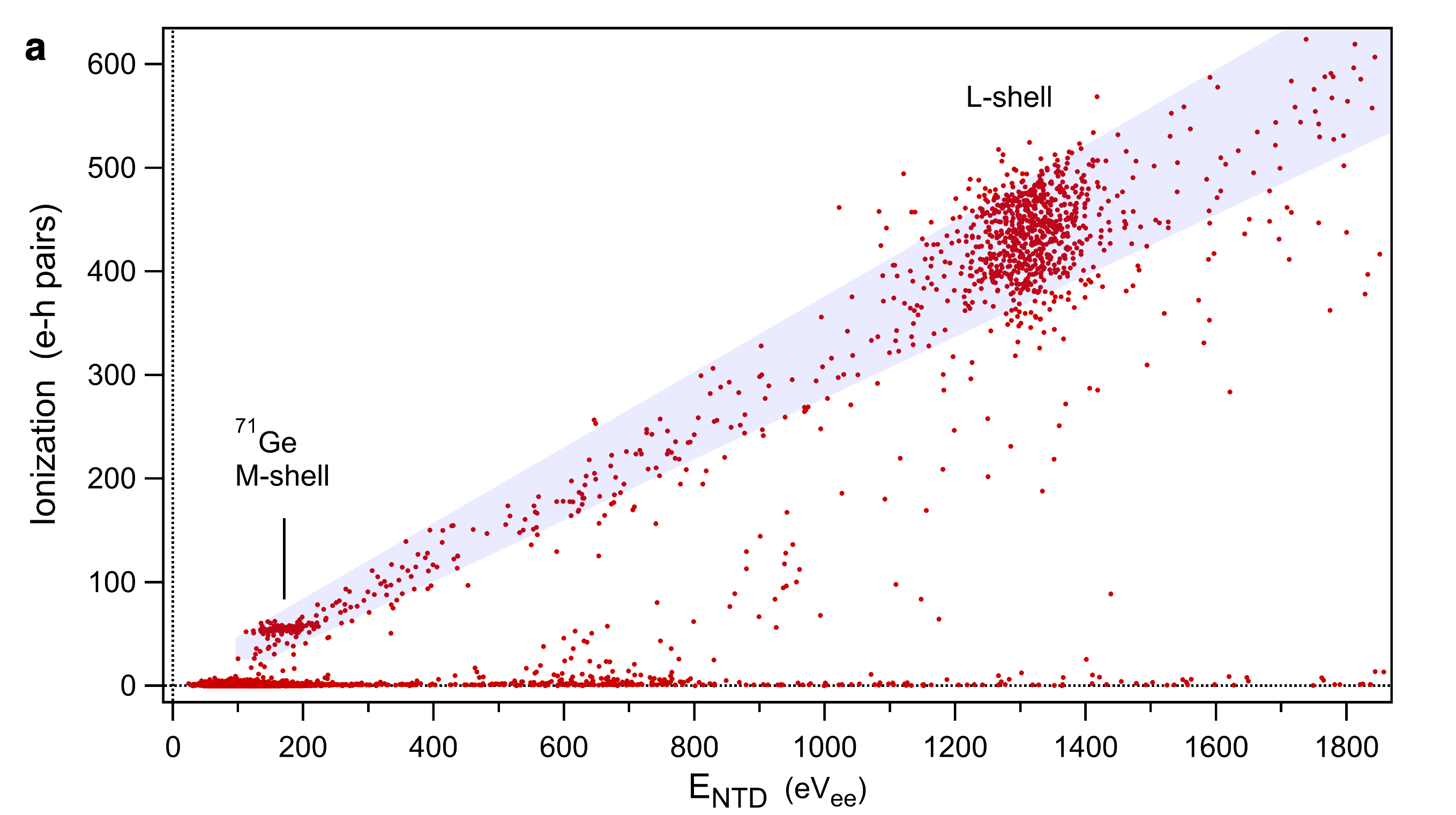}
  \includegraphics[width=\linewidth]{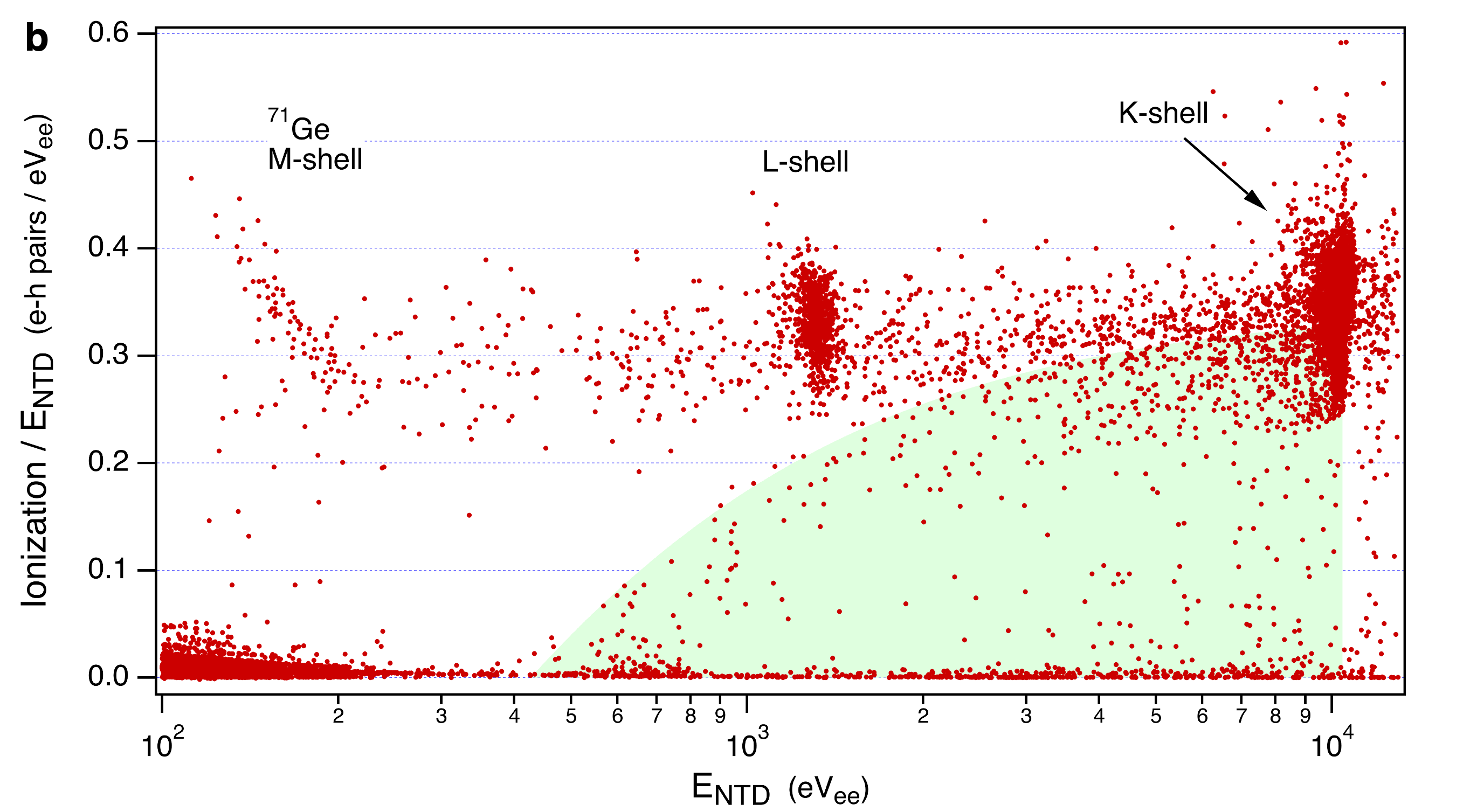}
     \caption{Dual-channel detection and particle discrimination.
     (a) Combined data corresponding to a total exposure of 11~h at an NTL voltage of 70~V. Events were triggered offline using the NTD-Ge channel. A well-defined electron-recoil band is observed (blue band), including the M- and L-shell emission lines from $^{71}$Ge. Two additional populations are visible: a non-ionizing background clustered near the $x$-axis, and an intermediate population between the electron-recoil and non-ionizing bands, attributed to charge trapping.
     (b) Ratio of TES-inferred ionization to the NTD-Ge signal as a function of NTD-Ge energy, expressed in eV$_\textrm{ee}$. Gamma events cluster around a ratio of $\sim$~1/3, whereas the non-ionizing background lies along the $x$-axis. Events with incomplete charge collection populate the intermediate region between these two bands; in particular, incomplete charge-collection K-shell events are expected to fall within the highlighted green band.}
  \label{fig:Quenching_plots}
\end{figure*}

Cryogenic semiconductor detectors enable active discrimination between events with different ionization quenching factors. This quantity is defined as the fraction of the deposited energy converted into ionization within the semiconductor and can be determined through simultaneous measurements of heat and ionization, combining calorimetric sensors with charge readout. Measurements of the ionization quenching factor allow the identification of nuclear recoils, electron recoils, and non-ionizing low-energy excess events, with the discrimination efficiency primarily governed by the device energy resolution \cite{Armatol:2025}.
As the detector threshold is approached, the baseline resolution of the charge amplifier, on the order of 10~e-h pairs rms in state-of-the-art Si and Ge devices, often becomes the dominant limitation, causing overlap between different event populations and reducing background discrimination. The single e-h pair sensitivity of our detector, combined with an accurate determination of the ionization efficiency (see Appendix~\ref{App_TESefficiency}), paves the way for robust rejection of low-energy non-ionizing backgrounds and, ultimately, discrimination between electron and nuclear recoils.  

This capability is illustrated in Fig.~\ref{fig:Quenching_plots}(a), which shows the TES-inferred ionization signal as a function of the total energy measured by the NTD-Ge thermistor, following the calibration procedure described in Appendix~\ref{App_Calibration}. Gamma-ray interactions, including the $^{71}$Ge K-, L- and M-shell lines, populate a well-defined diagonal band that is clearly separated from a cluster of non-ionizing events near the $x$-axis. This latter population is consistent with the low-energy excess observed in cryogenic detectors, which remains a major challenge for experiments requiring low-background spectroscopy at eV-scale energies.

The rejection performance is ultimately limited by the resolution of the phonon sensor, which is approximately 600~eV rms in phonon energy for the NTD-Ge thermistor used in the present detector. At an NTL bias of 70~V, phonon amplification during charge drift provides sufficient gain to reject the low-energy excess down to approximately 120~eV$_{\textrm{ee}}$. As shown in Fig.~\ref{fig:Quenching_plots}(a),(b), this enables a clear separation between the non-ionizing background and the $^{71}$Ge M-shell line. Ge detectors with crystal mass similar to our device have demonstrated phonon-energy resolutions more than an order of magnitude better using NTD-Ge thermistors \cite{Arnaud:2020}, while TES-based Si and sapphire cryogenic calorimeters routinely achieve baseline resolutions in the 1--10~eV rms range, nearly two orders of magnitude lower than that of the present detector \cite{Angloher:2024,Anthony-Petersen:2025}. Such performance would enable discrimination between electron recoils, nuclear recoils and non-ionizing background even for events approaching the single electron-hole-pair limit. A remaining limitation arises from non-ionizing interactions occurring near or within the TES, where enhanced phonon collection (see Appendix~\ref{App_TESefficiency}) can generate signals that mimic electron- or nuclear-recoil events.

By contrast, electron- or nuclear-recoil events affected by charge trapping exhibit a reduced TES-to-NTD-Ge signal ratio, causing them to fall below their respective recoil bands and toward, or into, the non-ionizing background population. Depending on the location of the trapped carriers and whether electrons or holes are trapped, part of the NTL energy is not released, either in the high- or low-electric-field regions of the crystal. The TES is particularly sensitive to a deficit in NTL phonons generated in the high-field region, producing the broad distribution of incomplete charge-collection events that extends across the region shown in Fig.~\ref{fig:Quenching_plots}(a),(b).

Events with reduced TES collection efficiency for NTL athermal phonons, such as those responsible for the low-energy tail observed in Fig.~\ref{fig:TES_eh_peaks}, also populate the intermediate region below the electron-recoil band. For $^{71}$Ge K-shell decays, the combined effects of incomplete charge collection and reduced TES efficiency shift events into the intermediate region highlighted by the green band in Fig.~\ref{fig:Quenching_plots}(b). Such events are clearly present in the data and can substantially degrade discrimination performance at low energies. The energy resolution of the thermal phonon sensor is therefore critical for efficiently rejecting low-energy excess events and incomplete charge-collection events that overlap with the expected electron- and nuclear-recoil signal bands.

Finally, the onset of leakage currents due to charge injection through the TES or Al Schottky barrier, observed at NTL biases $\gtrsim$~70~V, is expected to produce a distinct event population. As these events do not involve particle or photon interactions, they generate only NTL-induced athermal phonons, without accompanying recoil-energy phonons. In principle, they can be distinguished using the same discrimination scheme, appearing as a band slightly above the electron-recoil population in Fig.~\ref{fig:Quenching_plots}. Achieving robust separation between charge-injection and electron-recoil events, however, would require sub-eV phonon-sensor resolution, which remains extremely challenging in large-mass crystals.

\section{Discussion and outlook}

We have developed a high-purity germanium detector that combines a point-contact architecture with a low-temperature athermal-phonon sensor, enabling two key advances for low-energy spectroscopy. First, the point-contact geometry efficiently concentrates a substantial fraction of the energy released during charge drift onto a millimeter-scale TES layer, enhancing its sensitivity by more than two orders of magnitude. This enables single e-h pair ionization detection in a 40-g Ge crystal and establishes a pathway towards kilogram-scale high-purity Ge detectors with single-electron sensitivity. Second, this technology provides a robust approach for rejecting low-energy non-ionizing backgrounds and, ultimately, discriminating between electron and nuclear recoils. This capability relies on event-by-event measurements of the ionization quenching factor. Because the TES response is dominated by NTL athermal phonons, it provides a direct measurement of the ionization signal. Combined with an independent calorimetric sensor operating in thermal equilibrium, this enables simultaneous reconstruction of the recoil energy and ionization yield, allowing efficient separation of distinct event populations. Achieving single e-h sensitivity together with effective background discrimination represents a significant advance for sub-100-eV spectroscopy, with potential applications in neutron detection, coherent elastic neutrino-nucleus scattering (CE$\nu$NS), and light dark matter searches involving nuclear recoils.

The measured TES time response of approximately 10~$\mu$s supports counting rates up to $\sim$10~kHz when readout of the additional NTD-Ge thermal-phonon sensor is not required. Although slower than charge-amplifier-based devices with sub-microsecond intrinsic rise times, this response is well suited to coincidence-based experiments, where efficient vetoing is essential.

Fast monitoring of single-carrier drift further provides detailed insight into crystal properties and detector behavior, including single-charge-resolved leakage currents associated with the Schottky barrier at the Ge-electrode interface, charge trapping during drift, impact ionization, and tunnel ionization of residual Ge impurities. The temporal resolution of our device also enabled the identification of a dominant background of single e-h events at a rate of approximately 500 Hz, which we attribute to far-infrared stray radiation in the cryostat.

Further improvements in thermal-phonon sensor performance are expected to substantially enhance the recoil-energy resolution. This will not only strengthen background discrimination but also enable precise calibration of the Ge ionization quenching factor, particularly for nuclear recoils in the largely unexplored regime of 1--10~e-h pairs ($\lesssim$300~eV). Only a handful of experiments have measured the Ge ionization efficiency for nuclear recoils below 1~keV, and no data exist below 250~eV  \cite{Li:2025}. Experimental access to this energy range would provide a stringent test of Lindhard theory at very low energies \cite{Lindhard:1963,Lewin:1996} and deliver critical input for Ge-based dark matter direct-detection and CE$\nu$NS experiments. Finally, the detector design presented here is expected to yield similarly outstanding performance when implemented in high-purity silicon crystals.

\appendix
\section{Detector fabrication}\label{App_Fabrication}

\begin{figure*}[t]
  \centering
  \includegraphics[height=0.185\textheight]{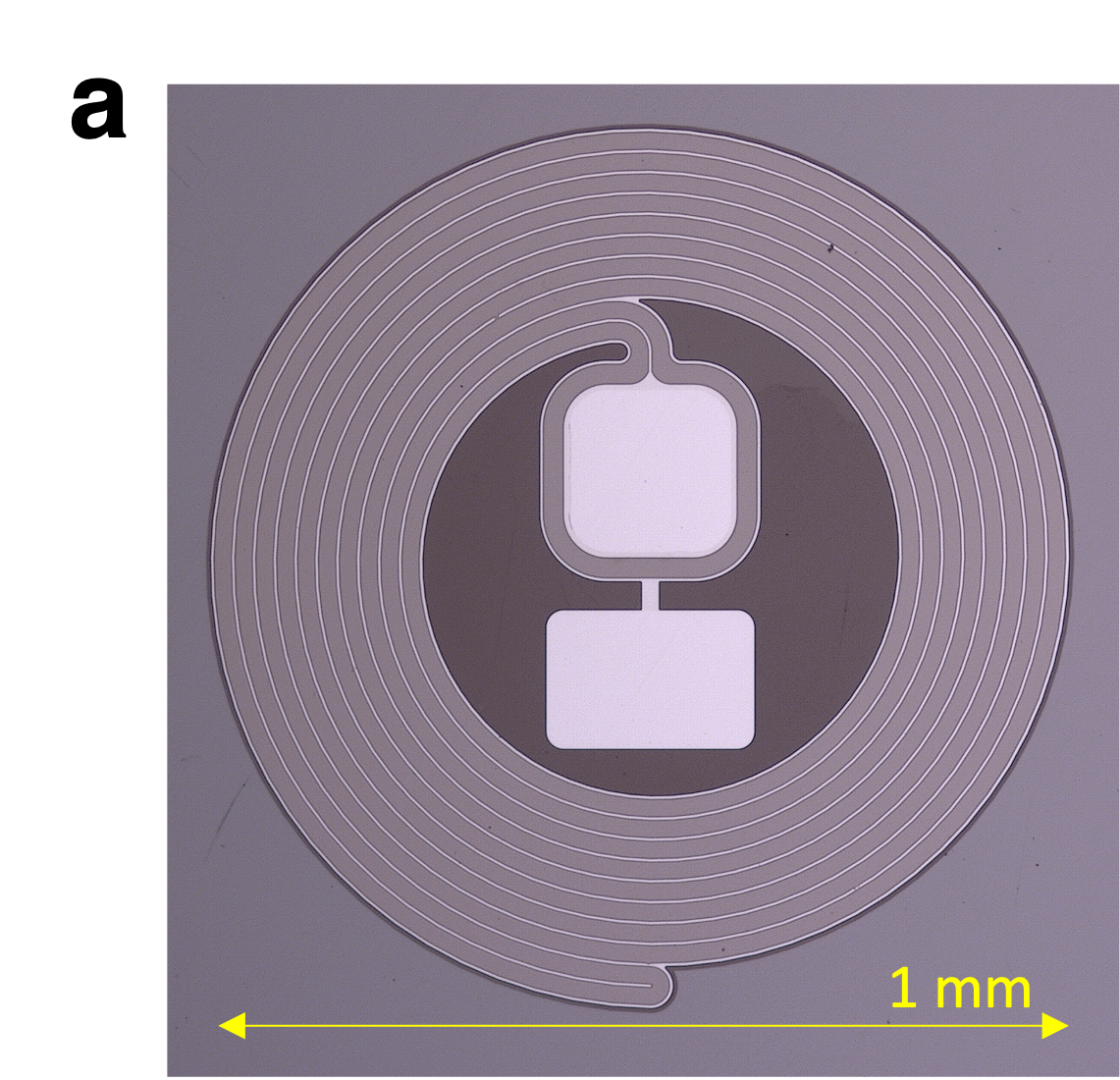}
  \includegraphics[height=0.178\textheight]{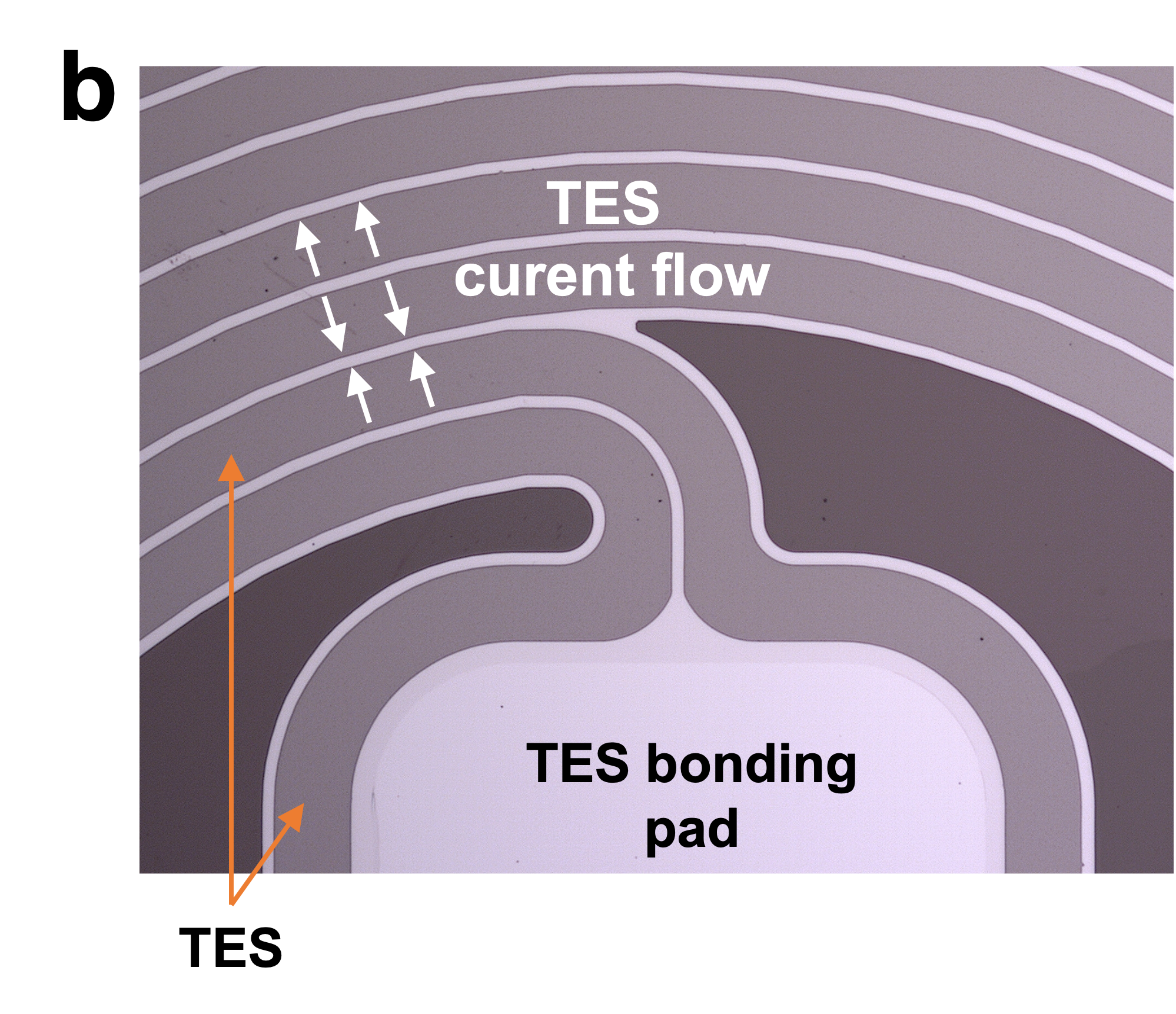}
  \includegraphics[height=0.18\textheight]{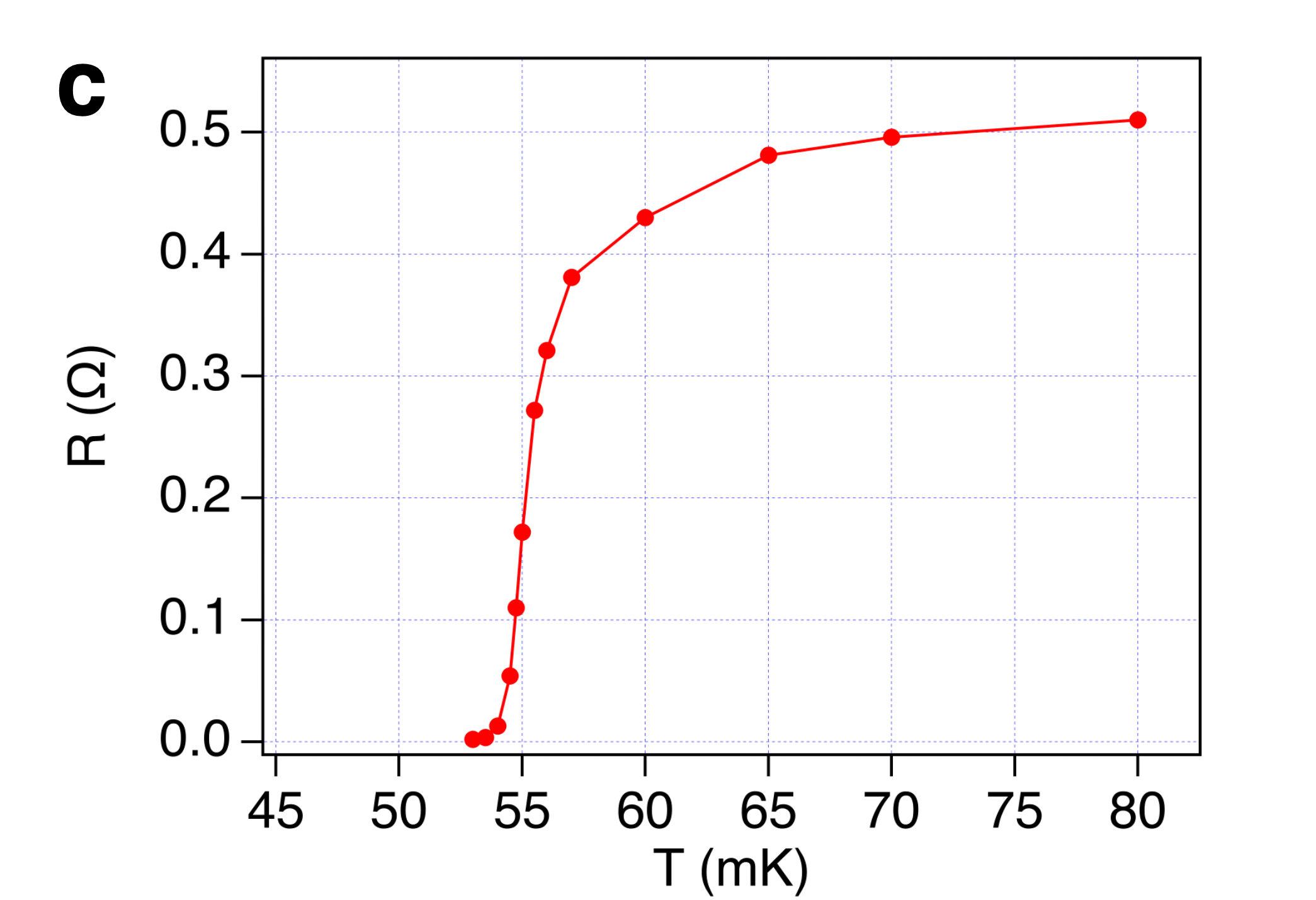}
  \includegraphics[height=0.223\textheight]{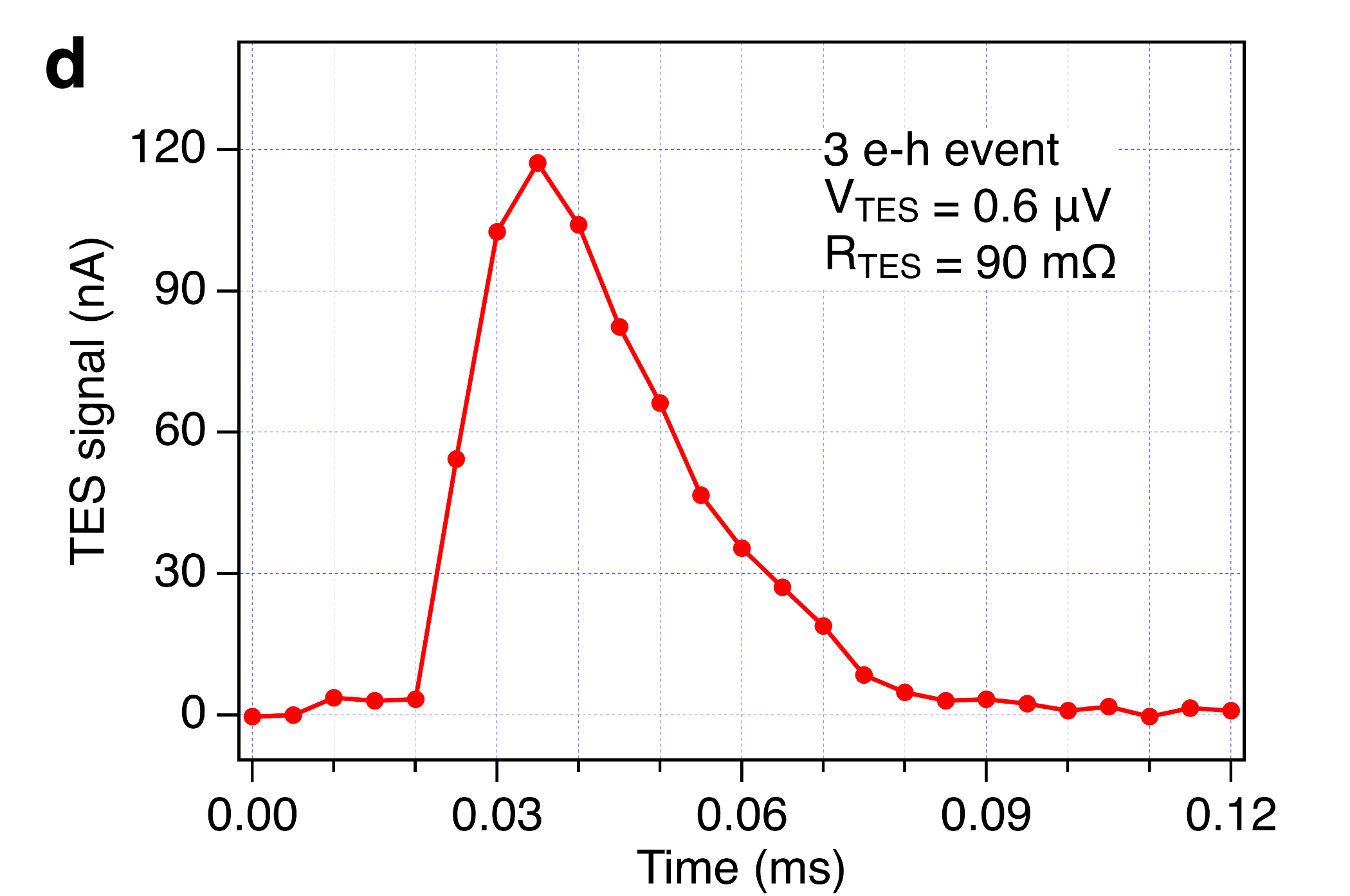}
  \includegraphics[height=0.223\textheight]{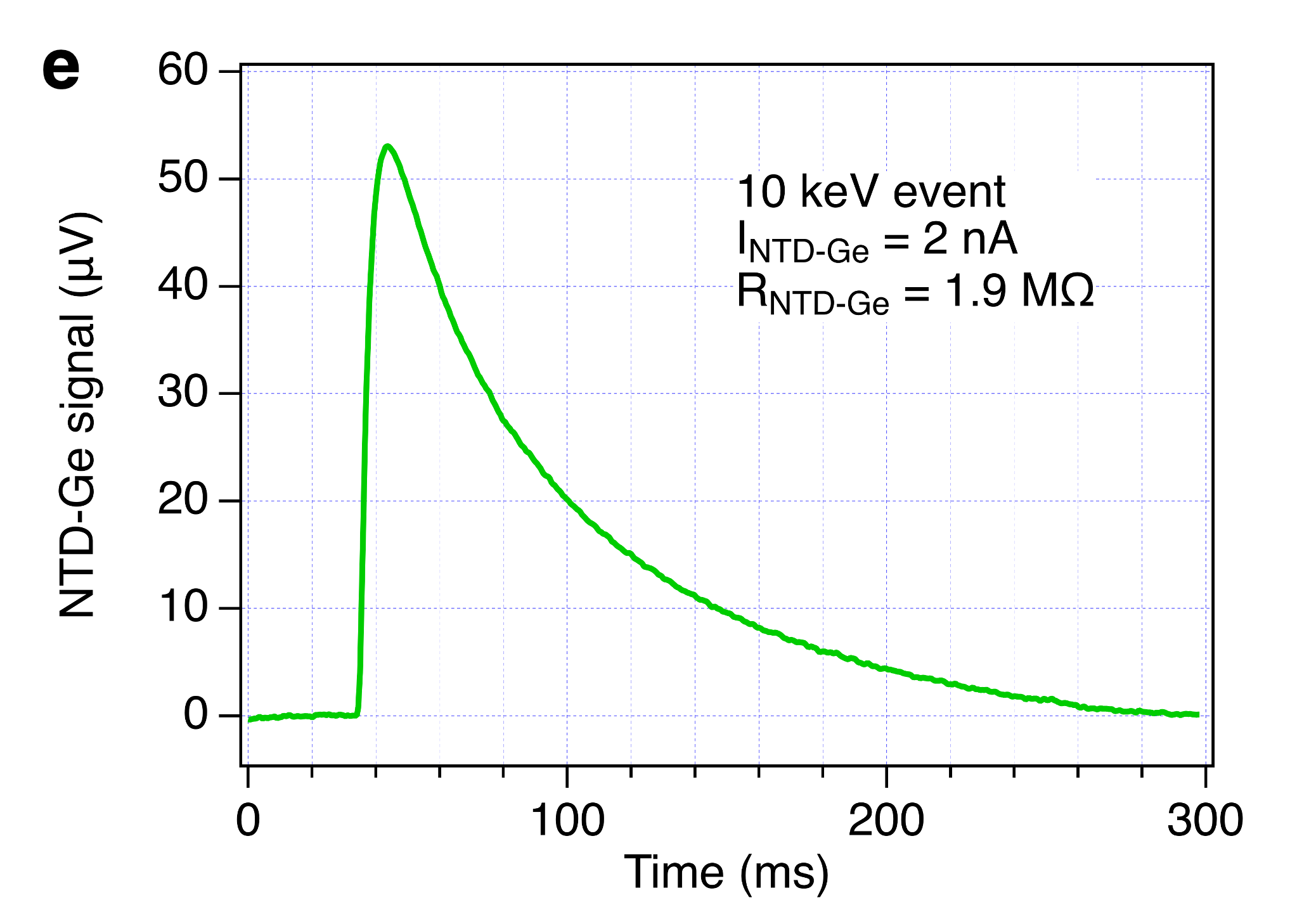}
    \caption{(a) Optical microscope image of the TES, biased via two Al electrodes patterned in a double-spiral geometry to achieve a high-aspect-ratio NbSi structure. The electrodes connect to central pads used for wedge bonding.
    (b) Close-up view of the sensor showing the NbSi layer (light grey), a bonding pad (white), and the TES biasing electrodes (white). The TES current flow between the electrodes is indicated by yellow arrows.
    (c) Superconducting transition curve of the TES measured at low bias.
    (d,~e) Representative pulses for a 3~e-h TES event and a 10.37~keV NTD-Ge thermistor event, measured at $V_\textrm{NTL}$~=~50~V under the biasing conditions used for data acquisition.}
  \label{fig:Methods_TES}
\end{figure*}

The detector is based on a commercial HPGe crystal (Mirion Technologies) with a residual impurity concentration $N_\textrm{A} - N_\textrm{D} < 1.5 \times 10^{10}$ cm$^{-3}$. The cylindrical crystal has a diameter of 30~mm, a height of 10~mm, and a mass of 40.5~g with rounded edges. It was delivered after mechanical polishing and chemical etching of its surfaces, and subsequently stored under a nitrogen atmosphere to minimize contamination and oxidation.

Detector fabrication was carried out at IJCLab and the C2N technological facility, and involved multiple stages of thin-film deposition and photolithography. The process begins with the deposition of a 30-nm-thick hydrogenated amorphous germanium (a-Ge:H) layer by electron-beam evaporation, covering the entire surface of the Ge crystal, including the top, bottom, and lateral faces. This layer reduces detector leakage current \cite{Looker:2015,Wei:2020} and enables operation at relatively high charge-collection voltages, thereby enhancing NTL energy release and improving the TES sensitivity. Prior in-situ cleaning of the crystal surface by Ar plasma ion-beam bombardment is essential to ensure strong adhesion of the a-Ge:H layer and to provide reliable wedge bonding of the Al electrode and the TES.

The TES  (Fig.~\ref{fig:Methods_TES}(a),(b)) is fabricated directly on top of the a-Ge:H layer by co-deposition of a 50-nm-thick Nb$_x$Si$_{1-x}$ film. The stoichiometry is tuned by adjusting the Nb and Si evaporation rates, with $x$~=~0.138 (Nb$_{0.138}$Si$_{0.862}$) yielding a superconducting transition temperature of approximately 55~mK (Fig.~\ref{fig:Methods_TES}(c)). Owing to its relatively high electrical sheet resistance \cite{Crauste:2013}, the NbSi film is patterned into a high-aspect-ratio geometry compatible with SQUID readout requirements. The sensor design features a width-to-length ratio of approximately 1100, resulting in a normal-state resistance of 0.55~$\Omega$. To protect the TES from oxidation and degradation during subsequent fabrication steps, a 30-nm-thick SiO layer is thermally evaporated on top of the sensor as a passivation coating.

The main detector electrode consists of a 150-nm-thick Al film deposited by electron-beam evaporation in multiple steps. It covers the bottom, lateral, and part of the top surface of the Ge crystal. On the top surface, the electrode is patterned into a large outer ring connected by a narrow line to a small field-shaping electrode surrounding the TES. To mitigate leakage currents in regions of high electric field, a 1-$\mu$m-thick SU-8 epoxy layer is embedded beneath the field-shaping-electrode. This highly insulating photoresist is deposited by spin-coating and patterned by photolithography prior to Al deposition. No SU-8 layer is placed beneath the NbSi TES in order to avoid scattering of NTL athermal phonons and to maximize their transmission from the Ge crystal to the sensor. As a result, a residual leakage current between the NbSi sensor and the Ge crystal remains possible and ultimately limits detector performance at high charge-collection voltages.

The field-shaping electrode provides two key advantages. It reduces charge trapping of drifting e-h pairs at the free Ge surface by redirecting electric-field lines into the crystal bulk, and it enhances TES sensitivity by increasing the local field near the sensor, thereby amplifying NTL phonon emission. Simulated electric-field lines and equipotential contours are shown in Fig.~\ref{fig:Detector}(a),(c). However, increasing the electric field in the vicinity of the TES also introduces drawbacks: above $\sim$10$^5$ V/cm, impact ionization during charge drift can trigger avalanche breakdown \cite{Miller:1955}, leading to permanent leakage currents that prevent proper detector operation.

The final fabrication step consists of gluing the NTD-Ge thermistor to the bottom surface of the crystal using a 10-$\mu$m-thick epoxy layer. This step is critical to ensure efficient thermal coupling to the Ge absorber while maintaining electrical isolation from the adjacent Al electrode. Any unintended electrical contact would introduce excess readout noise in the NTD-Ge channel and could damage the front-end electronics during operation at high collection voltages.

The completed detector is mounted in a Cu holder equipped with Kapton circuit boards for wedge bonding and cable soldering. The Ge crystal is supported by 2-mm-diameter sapphire balls on flexible blades to minimize mechanical stress during cooldown. Electrical connections to the NbSi TES and Al electrode are made by wedge bonding with 25-$\mu$m-diameter Al wires. Thermal coupling to the cryostat and electrical readout of the NTD-Ge thermistor are provided by 25-$\mu$m-diameter Au wires.

\section{Experimental apparatus}\label{App_ExpApparatus}

The detector was operated in a pulse-tube dilution refrigerator attaining a base temperature of 15~mK. The detector holder was mechanically decoupled from the cryostat structure by springs, reducing the transmission of pulse-tube vibrations and thus minimizing microphonic noise in the sensors \cite{Olivieri:2017}. Gamma-ray backgrounds and event pile-up in the NTD-Ge channel were reduced by partially enclosing the cryostat within an external shield composed of 10-cm-thick lead bricks. To minimize blackbody radiation backgrounds from the higher-temperature stages, the cryostat was equipped with screens attached to the 50~K, 4~K, 1~K and 15~mK flanges. The two coldest stages were coated with NEXTEL Velvet 811-21 to reduce stray photon reflections.

The detector readout chain employed a low-noise voltage amplifier for the NTD-Ge channel, based on a JFET front-end stage mounted inside the cryostat at 150~K. A two-stage STAR Cryoelectronics SQUID array system was used for the TES readout. During operation, a charge-collection voltage was applied to the Al electrode using a low-pass filtered DC generator, while the NbSi TES was held at ground potential through the SQUID electronics. The field-shaping-electrode surrounding the TES was biased at the same potential as the main Al electrode. The NTD-Ge thermistor was current-biased either in DC mode or in AC mode using a programmable square-wave generator. Data were synchronously acquired at a sampling rate of 200~kHz with a 16-bit NI USB-6366 multichannel ADC module. For each channel, a continuous data stream was recorded and analyzed offline. Low energy calibration and regeneration of the detector aimed at neutralizing space charge in the Ge crystal \cite{Olivieri:2009} were performed using a 1590-nm laser illuminating the detector through a single-mode optical fiber.

\section{Data analysis and calibration}\label{App_Calibration}

Simultaneous operation and optimization of the TES and NTD-Ge phonon sensors are essential for achieving a low energy threshold and robust background discrimination. Although both are low-temperature resistive sensors, their signals probe different aspects of the energy-deposition process in our point-contact device. The TES measures the non-equilibrium athermal phonon population generated following a particle interaction, whereas the NTD-Ge sensor measures the subsequent temperature increase after phonon down-conversion and thermalization. To maximize sensitivity to athermal phonons, the TES is voltage-biased in the strong electrothermal feedback regime and read out using a standard SQUID scheme \cite{Irwin:1995,Irwin:2005}. Under these conditions, the sensor self-stabilizes within its superconducting transition near 55~mK (Fig.~\ref{fig:Methods_TES}c), whereas the Ge absorber is maintained below 20~mK. The resulting temperature gradient substantially suppresses the TES response to fluctuations in the absorber base temperature. Operating the crystal below 20~mK also minimizes its heat capacity, thereby maximizing the sensitivity of the NTD-Ge thermistor.

TES data were analyzed using two complementary approaches: strong electrothermal-feedback self-calibration and pulse template fitting. Strong negative electrothermal feedback intrinsically stabilizes the TES temperature by dynamically compensating the absorbed athermal-phonon power through a corresponding reduction in Joule bias dissipation. As a result, the athermal phonon energy transmitted to the TES is efficiently extracted from the system and can be precisely measured via the TES current pulse:
\begin{equation}
    E_{\textrm{ETF}} =  \int V_\textrm{TES} \delta I(t) ~dt.
    \label{Eq_integralETF}
\end{equation}
Here, $E_\textrm{ETF}$ is the electrothermal feedback energy extracted from the TES, $V_\textrm{TES}$ is the TES voltage bias, assumed constant in this work, and $\delta I(t)$ is the current pulse measured by the SQUID amplifier. In cases where $V_\textrm{TES}$ varies during the pulse, second-order corrections are required. The integral is calculated over the pulse duration.

The dynamic range of this self-calibration method is primarily determined by the applied TES bias voltage, which limits the available Joule heating power. At low bias voltages, the athermal phonon energy transmitted to the TES electrons cannot be evacuated efficiently, causing a fraction of the energy to be re-emitted into the Ge crystal via electron-phonon coupling in the NbSi layer. A similar effect occurs at sufficiently high deposited energies, even for large TES bias voltages; in our device, this regime is reached at approximately 100~e-h pairs for an NTL collection bias of 50~V. In both cases, the TES response becomes nonlinear. 

Despite the large volume of the Ge crystal, the TES pulses are remarkably fast, exhibiting rise and decay times approximately three orders of magnitude shorter than those of the NTD-Ge pulses (Fig.~\ref{fig:Methods_TES}(d),(e)). The TES decay time depends strongly on the applied bias voltage as a result of negative electrothermal feedback. At low bias, where this feedback is negligible, the 90--30\% decay time is of order 1 ms and is primarily governed by electron-phonon coupling in the NbSi layer  \cite{Marnieros:2012}. This timescale corresponds to the relaxation process through which athermal phonon energy is transferred from the TES back to the Ge crystal. For the calibration runs presented here, the TES was biased at 0.6~$\mu$V and $R=$~90~m$\Omega$, yielding a much shorter decay time of approximately 20~$\mu$s for low-energy events. This timescale corresponds to the evacuation of the athermal phonon energy trapped in the TES via electrothermal feedback. For energies above 1~keV, the linear dynamic range of the TES and SQUID readout chain is exceeded, leading to signal distortion and longer decay times. The pulse rise time (10--90\%) is approximately 10~$\mu$s and shows little dependence on TES bias. Owing to the point-contact geometry, the rise time of ionizing events is largely independent of interaction position, as charge carriers predominantly generate NTL athermal phonons in the high-field region beneath the TES. The resulting TES response is therefore fast compared with conventional low-temperature calorimeters, enabling operation at counting rates exceeding several kHz with minimal dead time and pile-up of events. By contrast, the NTD-Ge sensor measures the temperature increase of the detector after phonon thermalization and produces substantially slower pulses, with typical rise and decay times of approximately 2~ms and 60~ms, respectively. 

For low-energy TES data, below approximately 10~e-h pairs, a second analysis method based on time-domain pulse template fitting was also employed. A TES pulse template is constructed by averaging several hundred low-energy laser-induced events. Each TES pulse is then fitted with this template, yielding the pulse amplitude and mean squared error, along with auxiliary parameters such as baseline level and slope, and rise and decay times. The fit is performed over a 200~$\mu$s trace extracted from the digitized data stream and centered on the trigger position. Triggers are provided either by the laser electronics or by a peak-detection algorithm that scans the recorded data stream for maxima above a fixed threshold after application of a first-order Butterworth band-pass filter (2--50~kHz). Although the template-fitting method achieves a lower TES 5$\sigma$ energy threshold and improved baseline resolution, it is less effective over a wide energy range owing to the non-homothetic pulse shapes: higher-energy pulses exhibit longer decay times that cannot be accurately fitted by a single template.

In contrast, NTD-Ge signals can be reliably described by a single pulse template over an extended energy range. The NTD-Ge data considered in this work were therefore analyzed exclusively using the time-domain template-fitting algorithm and calibrated with the $^{71}$Ge K-, L- and M-shell emission lines. As discussed above, the NTD-Ge sensor measures the temperature rise of the Ge crystal following down-conversion of athermal phonons. Because the NTD-Ge rise time is significantly longer than the TES decay time, the electrothermal feedback energy $E_{\textrm{ETF}}$ extracted by the TES modifies the NTD-Ge response and must be accounted for in the calibration. The total energy measured by the NTD-Ge is therefore
\begin{equation}
     E_{\textrm{NTD}}=E_{\textrm{R}}+E_{\textrm{NTL}}-E_{\textrm{ETF}},
    \label{Eq_TotalEnergyNTD}
\end{equation}
where $E_{\textrm{R}}$ is the recoil energy deposited locally by the interacting particle or photon, and $E_{\textrm{NTL}}$ the NTL energy released as excited e-h pairs drift through the crystal.

The NTD-Ge energy spectrum is commonly expressed in electron-volts electron-recoil equivalent (eV$_\textrm{ee}$), assuming interactions deposit energy via electron recoils with an ionization yield of one e-h pair per 3~eV in Ge. Under this assumption, the NTL phonon contribution is given by $E_{\textrm{NTL}}=E_{\textrm{R}}/3\cdot qV_\textrm{NTL}$, and Eq.~\eqref{Eq_TotalEnergyNTD} becomes
\begin{equation}
    E_{\textrm{NTD}}=E_{\textrm{R}}(1+\frac{qV_\textrm{NTL}}{3})-E_{\textrm{ETF}}.
    \label{Eq_TotalEnergyNTD_eVee}
\end{equation}
where $q$ is the elementary charge. When energies are expressed in electron-volts, $q=1$ and $qV_\textrm{NTL}$ is numerically equal to the applied NTL voltage expressed in volts.

Calibration of the NTD-Ge signal is achieved by aligning the centroids of Gaussian fits to the $^{71}$Ge K-, L- and M-shell X-ray peaks with energies calculated using Eq.~\eqref{Eq_TotalEnergyNTD_eVee} with $E_{\textrm{R}}=$ 10.37~keV, 1.3~keV and 160~eV, respectively. The corresponding $E_{\textrm{ETF}}$ values are determined from the TES signal using Eq.~\eqref{Eq_integralETF}.

\section{TES signal model and phonon collection efficiency}\label{App_TESefficiency}

\begin{figure*}[!ht]
  \centering
  \includegraphics[width=\linewidth]{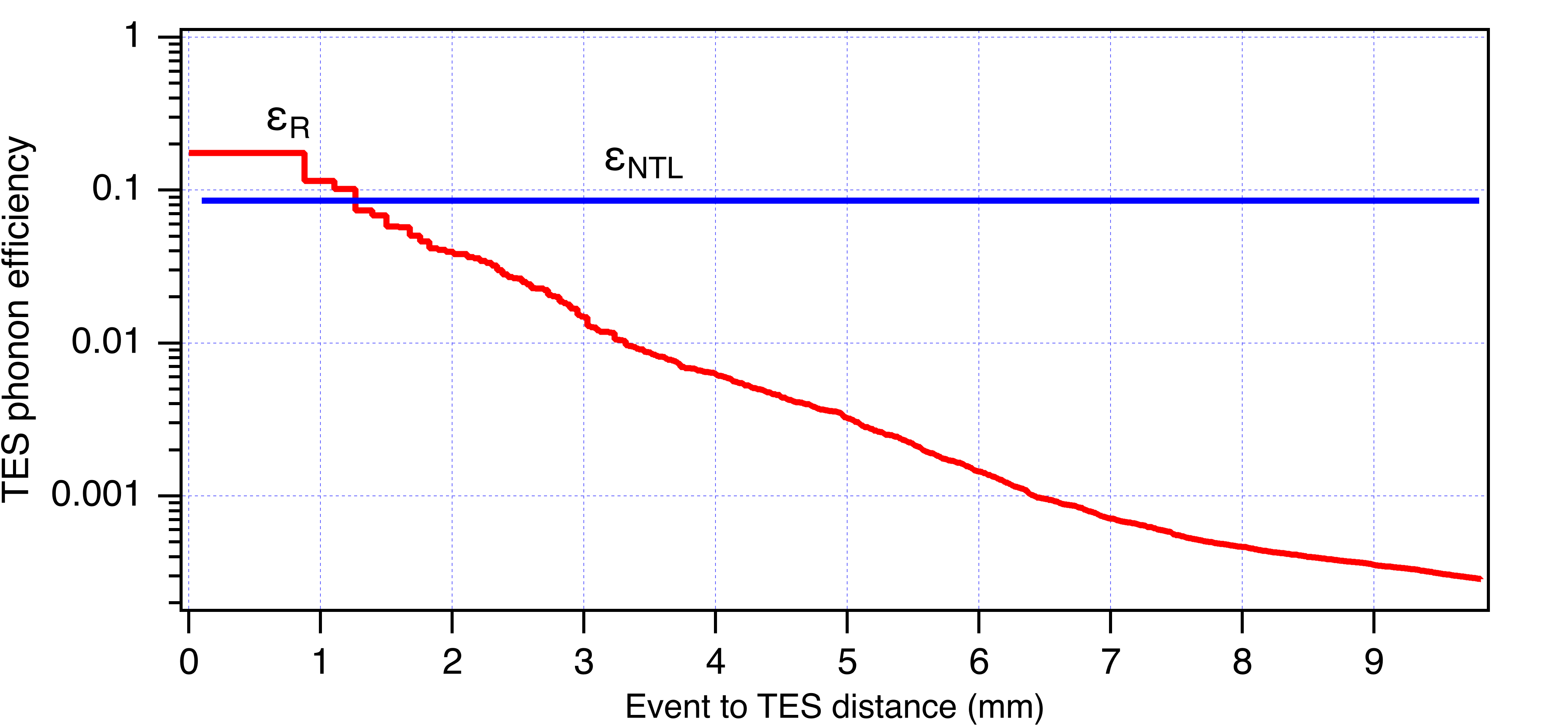}
     \caption{Estimated TES efficiency for collecting athermal phonons generated by 10.37~keV recoils (red line) and NTL e-h drift (blue line) as a function of event localization. The recoil phonon efficiency is calibrated using data acquired at zero NTL collection bias.}
  \label{fig:Methods_TES_epsilonR}
\end{figure*}

Unlike the NTD-Ge sensor, which exhibits comparable sensitivity to both $E_{\textrm{R}}$ and $E_{\textrm{NTL}}$ contributions, the point-contact TES is significantly more sensitive to NTL energy release. This enhanced response arises from strong coupling to NTL phonons emitted in the high-field region of the crystal and, owing to the point-contact design, is largely independent of interaction position. By contrast, athermal phonons associated with recoil energy deposited directly in the Ge lattice propagate through the crystal and are absorbed by the Al electrode and the TES with a strongly position-dependent efficiency. The TES response is enhanced for events occurring near the sensor but decreases rapidly with distance, becoming negligible beyond a few millimeters. This behavior results from the extensive Al coverage of the Ge surface, which traps athermal phonons with high efficiency.

The total athermal phonon energy trapped in the TES can be expressed by the following equation:

\begin{equation}
    E_\textrm{TES} = \varepsilon_{\textrm{R}} E_{\textrm{R}} + \varepsilon_\textrm{NTL} E_\textrm{NTL},
    \label{Eq_ETES}
\end{equation}
where $\varepsilon_{\textrm{R}}$ is the TES recoil-phonon efficiency, describing the fraction of the recoil energy transferred to the sensor via athermal phonons and $\varepsilon_\textrm{NTL}$ is the TES NTL-phonon efficiency.

Operating the detector at zero collecting bias suppresses the NTL contribution, enabling a direct determination of $\varepsilon_{\textrm{R}}$ and its dependence on the event position. In our device, the calibration of $\varepsilon_{R}$ is based on the 10.3~keV X-ray events from $^{71}$Ge, which are uniformly distributed throughout the crystal. The NTD-Ge thermistor baseline resolution allows efficient identification of these events above the background. Events within $\pm3\sigma$ of the 10.3~keV peak are selected, and their $E_\textrm{ETF}$ values are calculated following Eq.~\eqref{Eq_integralETF}. Assuming a simplified model in which the TES response decreases monotonically with the distance $d$ between the interaction point and the sensor, $\varepsilon_{\textrm{R}}(d)$ is extracted from the measured $E_\textrm{ETF}$ spectrum. Within this model, the Ge crystal is divided into $N$ equal-volume elements, where $N$ corresponds to the number of selected $^{71}$Ge K-shell events (approximately $10^{4}$). Each element is assigned a different distance from the TES, ranging from events occurring close to the sensor to those farther away. Since the true interaction position is not directly measured, we use the signal amplitude as a proxy for the event-to-TES distance. This approximation relies on the assumption that events occurring closer to the TES produce larger signals owing to more efficient phonon collection. The events are therefore sorted by decreasing amplitude, which approximately corresponds to increasing distance from the TES. For each event, we calculate the corresponding $\varepsilon_{\textrm{R}}$ and associate it with the inferred event-to-TES distance (Fig.~\ref{fig:Methods_TES_epsilonR}).
We find $\varepsilon_{\textrm{R}}\approx 0.2$ for events occurring within 1~mm of the TES, whereas athermal phonon trapping in the Al electrode dominates at larger distances, causing $\varepsilon_{\textrm{R}}$ to decrease rapidly to values below $10^{-3}$.

The TES collection efficiency for NTL phonons, $\varepsilon_\textrm{NTL}$, is calibrated at finite Ge bias in the range 15--90~V using low-energy laser events. As shown below, under these conditions the recoil-energy contribution $\varepsilon_{\textrm{R}} E_{\textrm{R}}$ in Eq.~\eqref{Eq_ETES} can be neglected. The peak positions of the $E_\textrm{ETF}$ spectrum from laser events, divided by the corresponding NTL energy, $E_{NTL}=Q\cdot V_\textrm{NTL}$, provide a direct measurement of $\varepsilon_\textrm{NTL}$, where $Q$ is the number of excited e-h pairs. The measured $\varepsilon_\textrm{NTL}$ decreases slowly with increasing bias, from 0.095 at 15~V to 0.08 at 90~V. These values enable the TES to reach a $5\sigma$ single e-h threshold for Ge biases above approximately 20~V. The observed decrease of $\varepsilon_\textrm{NTL}$ with bias is attributed to charge accumulation beneath the SU-8 sub-layer of the field-shaping electrode, originating from leakage currents at high voltage during bias ramp-up. These charges partially screen the  field-shaping electrode potential, modifying the electric field configuration in the crystal and reducing the effective potential drop and the NTL phonon emission in the vicinity of the TES. 

In the NTL bias range relevant for achieving low energy thresholds in our device (typically 30--90~V), the recoil-energy contribution to the TES signal is significantly smaller than the NTL component for ionizing events and can generally be neglected. For example, for gamma interactions with an ionization yield of one e-h pair per 3~eV and an NTL bias of 60~V, the ratio of recoil to NTL energy is
\begin{equation}
    \frac{E_{\textrm{R}}}{ E_\textrm{NTL}}  = \frac{E_{\textrm{R}}}{(Q \cdot V_\textrm{NTL})}  = \frac{E_{\textrm{R}}}{(E_{\textrm{R}}/3 \cdot V_\textrm{NTL})}  = \frac{1}{20}.
\end{equation}

Moreover, the TES collection efficiency for recoil-induced athermal phonons decreases strongly for events occurring more than  $\sim$1.5~mm from the sensor (Fig.~\ref{fig:Methods_TES_epsilonR}),  further suppressing the ratio $\varepsilon_{\textrm{R}} E_{\textrm{R}} / \varepsilon_\textrm{NTL} E_\textrm{NTL}$  for most gamma interactions. Although nuclear recoils are less ionizing, their contribution to the recoil-energy signal remains subdominant. Instead, the laser photon events used to calibrate $\varepsilon_\textrm{NTL}$ are more ionizing than gamma interactions and produce a negligible recoil energy contribution, validating the initial assumption.
Under these conditions, the TES signal provides a direct probe of the ionization yield in the Ge crystal and can be approximated as
 
\begin{equation}
E_{TES} \approx \varepsilon_\textrm{NTL} \cdot E_\textrm{NTL} = \varepsilon_\textrm{NTL}\cdot Q \cdot V_\textrm{NTL}.
\end{equation}

Combining a point-contact TES, which primarily measures $E_\textrm{NTL}$ with a low-threshold thermistor that responds to both $E_\textrm{NTL}$ and $E_{\textrm{R}}$, enables an independent determination of the ionization efficiency $Q/E_{\textrm{R}}$ and provides a pathway towards discrimination between electron recoils, nuclear recoils, and non-ionizing low-energy events.

Finally, complementary measurements using the K-, L- and M-shell X-ray interactions of  $^{71}$Ge extend the TES calibration to higher energies beyond the sensor’s linear dynamic range.

\section{Space-charge neutralization}\label{App_SpaceCharge}

\begin{figure*}[!ht]
  \centering
  \includegraphics[height=0.24\textheight]{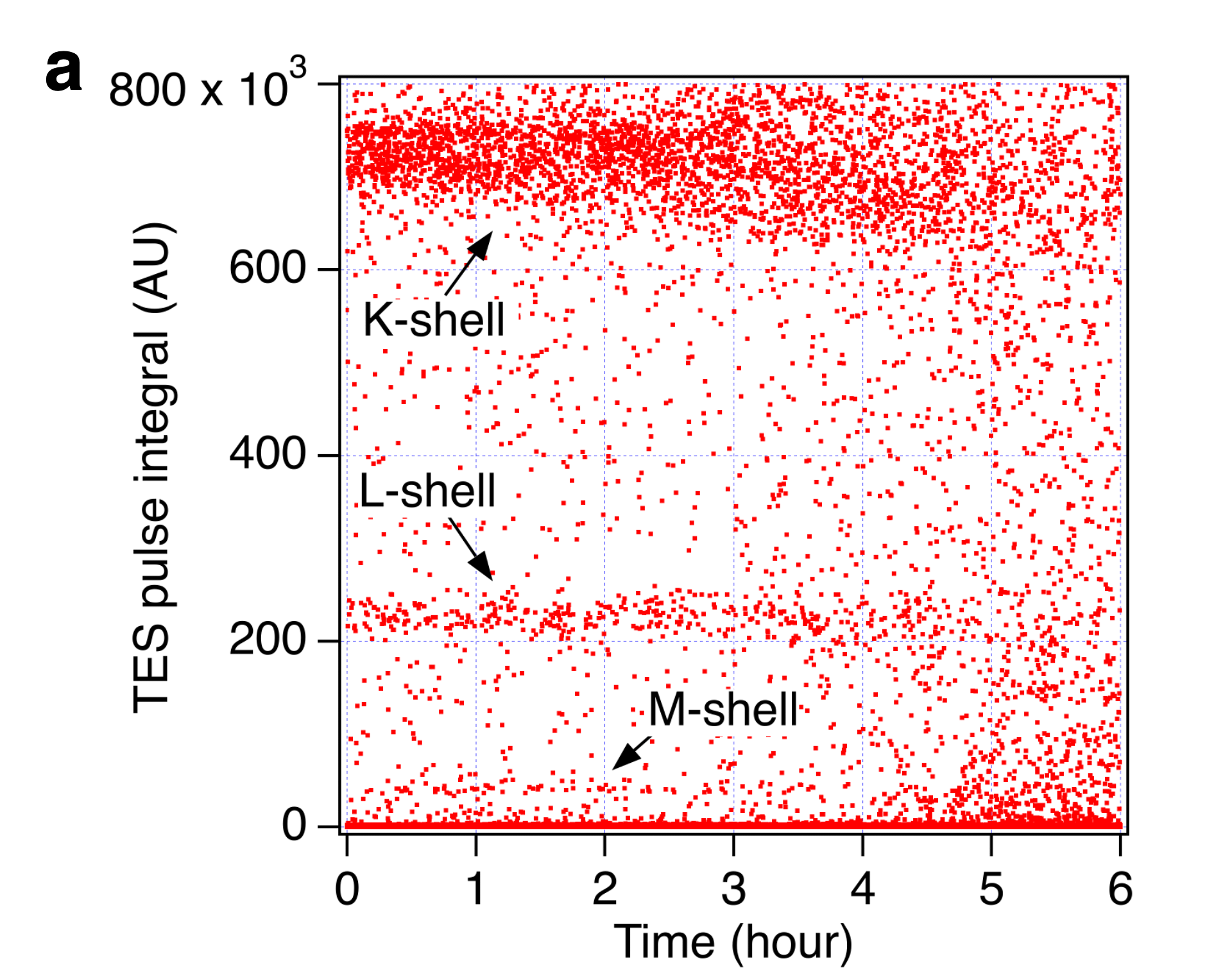}
  \includegraphics[height=0.24\textheight]{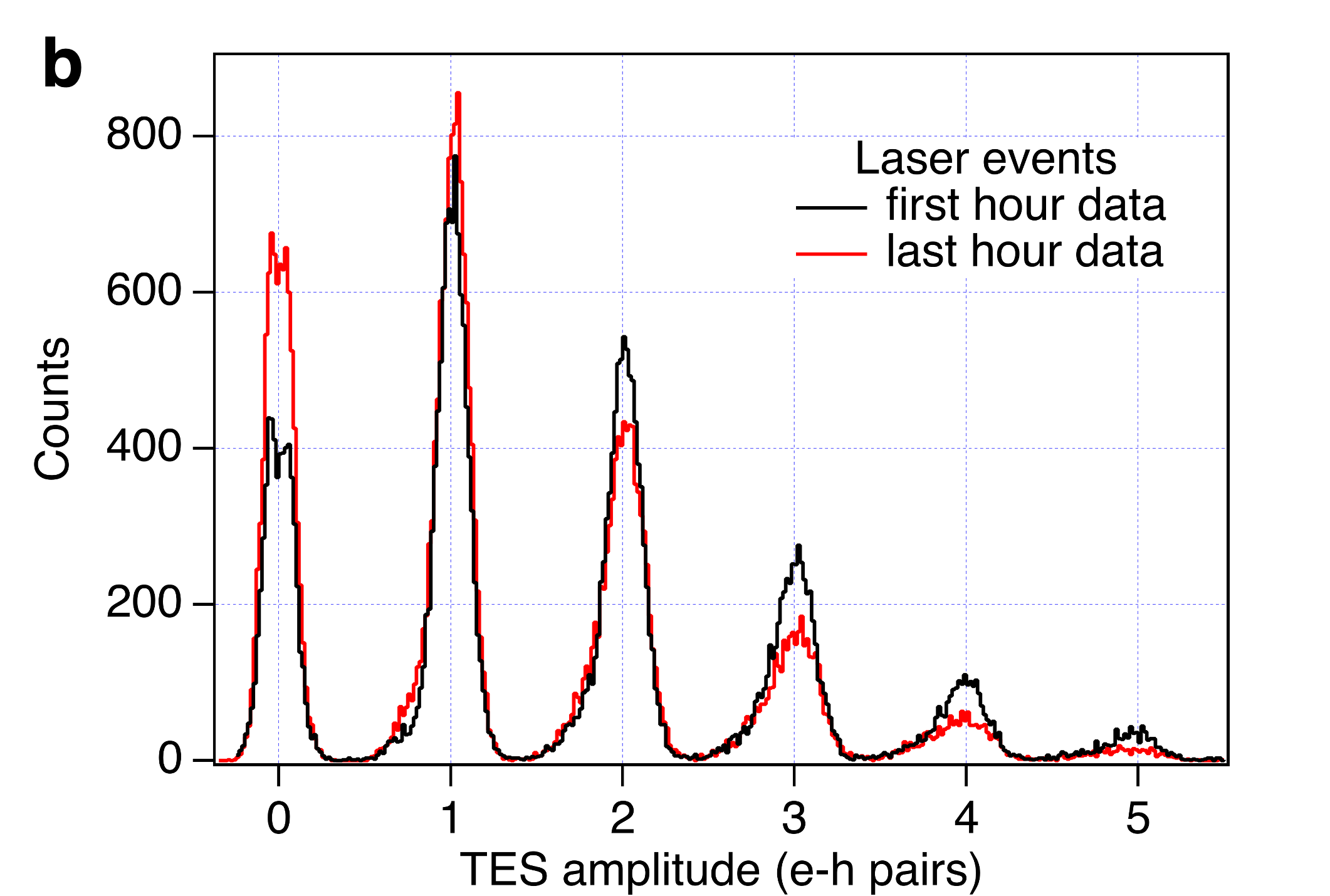}
     \caption{Degradation of charge collection during a 6-h run at 50 V.
     \textbf{a,}~Integral of TES pulses showing progressive degradation due to charge trapping in the $^{71}$Ge K-, L- and M-shell events.
     \textbf{b,}~Evolution of the TES spectra for low-energy laser events (0–5 e-h excitations), comparing the first and last hours of the 6-h acquisition.}
  \label{fig:Methods_SpaceCharge}
\end{figure*}

Activation of the Ge crystal with a $^{252}$Cf neutron source prior to cool-down enables monitoring of the detector charge-collection efficiency through the observation of the $^{71}$Ge K-, L-, and M-shell X-ray emission lines. The $^{71}$Ge isotope is produced homogeneously throughout the crystal by neutron irradiation and has a half-life of 11.5~days \cite{Norman:2024}, resulting in an approximately constant emission rate over the duration of our few-hour acquisition runs.

Figure \ref{fig:Methods_SpaceCharge} shows the evolution of TES signals during a 6 h run at a collecting voltage of 50~V. During the first three hours, the K-, L-, and M-shell event rates remain stable, indicating high and steady charge-collection efficiency. In the final three hours, however, significant space charge accumulates within the Ge crystal, progressively degrading detector performance and leading to broadening and partial loss of the $^{71}$Ge X-ray emission lines.

The space charge originates from the ionization of residual bulk and surface impurities in the Ge crystal \cite{Shutt:1992,LHote:2000,Broniatowski:2006}. These ionized impurities screen the applied electric field, reducing both charge-collection efficiency and energy resolution. Impurity ionization results from carrier trapping during charge drift as well as from stray infrared radiation incident on the crystal. To neutralize the accumulated space charge and restore nominal detector performance, a regeneration procedure is performed after several hours of data acquisition. During this process, the NTL bias is grounded and the laser is operated at high photon flux, temporarily raising the detector base temperature above 100~mK. A typical regeneration cycle lasts approximately 30 minutes, consisting of 20 minutes of laser-induced space-charge neutralization followed by 10 minutes for thermal recovery of the detector base temperature.

Space-charge degradation predominantly affects the free surface at the top side of the Ge crystal and the low-electric-field regions near its lateral surfaces, whereas the high-field region beneath the TES remains comparatively resilient. This behavior is evident in the low-energy laser calibration data shown in Fig.~\ref{fig:Methods_SpaceCharge}(b), taken from the same run as Fig.~\ref{fig:Methods_SpaceCharge}(a). Despite the progressive degradation of the  $^{71}$Ge X-ray emission lines, the positions of the single e-h laser peaks remain stable throughout the run. This behavior reflects the strong impact of space charge on carrier transport in low-field regions, while charge drift in the high-field region beneath the TES, where the majority of NTL athermal phonons are generated, remains largely unaffected. Consequently, charge trapping shifts the Poisson-distributed laser spectrum towards lower mean occupancy without significantly altering the positions of the individual e-h peaks. Over the final hour of data taking, the mean number of generated pairs per laser pulse decreased from $\mu=1.72$ to $\mu=1.36$ owing to charge trapping. The buildup of space charge is likely driven by high-energy cosmic-ray interactions in the crystal, which generate trapped charge carriers. Operation in an underground environment is therefore expected to substantially suppress space-charge accumulation and extend detector operating times between regeneration cycles.
Potential drifts in laser wavelength or intensity would likewise modify the mean occupancy of the charge spectrum but are not expected to affect the positions of the individual e-h peaks.

\begin{acknowledgments}

This work was supported by the French National Research Agency (ANR) through the CRYOSEL project (ANR-21-CE31-0004), by the France 2030 program through the TES4DM project (ANR-24-RRII-0001), and was partially supported by the French RENATECH network.
Y.Q. and W. G. acknowledge support provided by the National High Magnetic Field Laboratory at Florida State University, which is supported by the National Science Foundation Cooperative Agreement No. DMR-2128556 and the state of Florida.
\end{acknowledgments}

\section*{Data availability}
The data that support the findings of this study are available from the corresponding author upon reasonable request.

\bibliography{sn-bibliography}

@article{Acharya:2023,
  author       = {Acharya, P. and others},
  title        = {{Observation of time-dependent internal charge amplification in a planar germanium detector at cryogenic temperature}},
  journal      = {Eur. Phys. J. C},
  volume       = {83},
  number       = {278},
  year         = {2023},
  url          = {https://doi.org/10.1140/epjc/s10052-023-11432-y}
}

@article{Ackermann:2025,
  author       = {N. Ackermann and others},
  title        = {{Direct observation of coherent elastic antineutrino–nucleus scattering}},
  journal      = {Nature},
  volume       = {643},
  number       = {8074},
  pages        = {1229--1233},
  year         = {2025},
  url          = {https://doi.org/10.1038/s41586-025-09322-2}
}

@article{Ackermann:2024,
  author       = {N. Ackermann and others},
  collaboration = {CONUS Collaboration},
  title        = {{CONUS$^{+}$ Experiment}},
  journal      = {Eur. Phys. J. C},
  volume       = {84},
  number       = {12},
  pages        = {1265},
  year         = {2024},
  url          = {https://doi.org/10.1140/epjc/s10052-024-13551-6}
}

@article{Adamski:2025,
  title = {{Evidence of Coherent Elastic Neutrino-Nucleus Scattering with COHERENT's Germanium Array}},
  author = {Adamski, S. and others},
  collaboration = {The COHERENT Collaboration},
  journal = {Phys. Rev. Lett.},
  volume = {134},
  issue = {23},
  pages = {231801},
  numpages = {7},
  year = {2025},
  month = {Jun},
  publisher = {American Physical Society},
  url = {https://doi.org/10.1103/PhysRevLett.134.231801}
}

@Article{Adari:2022,
	title={{EXCESS workshop: Descriptions of rising low-energy spectra}},
	author={P. Adari and others},
	journal={SciPost Phys. Proc.},
	pages={001},
	year={2022},
	publisher={SciPost},
	url={https://doi.org/10.21468/SciPostPhysProc.9.001}
}

@article{Aggarwal:2025,
  title = {{Probing Benchmark Models of Hidden-Sector Dark Matter with DAMIC-M}},
  author = {Aggarwal, K. and others},
  collaboration = {DAMIC-M Collaboration},
  journal = {Phys. Rev. Lett.},
  volume = {135},
  issue = {7},
  pages = {071002},
  numpages = {10},
  year = {2025},
  month = {Aug},
  publisher = {American Physical Society},
  url = {https://doi.org/10.1103/2tcc-bqck}
}

@article{Agnese:2018,
  title = {{First Dark Matter Constraints from a SuperCDMS Single-Charge Sensitive Detector}},
  author = {Agnese, R. and others},
  collaboration = {SuperCDMS Collaboration},
  journal = {Phys. Rev. Lett.},
  volume = {121},
  issue = {5},
  pages = {051301},
  numpages = {7},
  year = {2018},
  month = {Aug},
  publisher = {American Physical Society},
  url = {https://doi.org/10.1103/PhysRevLett.121.051301}
}

@article{Albakry:2025,
  title = {{Light dark matter constraints from SuperCDMS HVeV detectors operated underground with an anticoincidence event selection}},
  author = {Albakry, M. F. and others},
  collaboration = {SuperCDMS Collaboration},
  journal = {Phys. Rev. D},
  volume = {111},
  issue = {1},
  pages = {012006},
  numpages = {9},
  year = {2025},
  month = {Jan},
  publisher = {American Physical Society},
  url = {https://doi.org/10.1103/PhysRevD.111.012006}
}

@article{Albakry:2026,
  title = {{Search for low-mass electron-recoil dark matter using a single-charge sensitive SuperCDMS-HVeV detector}},
  author = {Albakry, M. F. and others},
  collaboration = {SuperCDMS Collaboration},
  journal = {Phys. Rev. D},
  volume = {113},
  issue = {3},
  pages = {032001},
  numpages = {10},
  year = {2026},
  month = {Feb},
  publisher = {American Physical Society},
  url = {https://link.aps.org/doi/10.1103/5lnp-6mng}
}

@article{Angloher:2024,
  title = {{First observation of single photons in a CRESST detector and new dark matter exclusion limits}},
  author = {Angloher, G. and others},
  collaboration = {CRESST Collaboration},
  journal = {Phys. Rev. D},
  volume = {110},
  issue = {8},
  pages = {083038},
  numpages = {11},
  year = {2024},
  month = {Oct},
  publisher = {American Physical Society},
  url = {https://link.aps.org/doi/10.1103/PhysRevD.110.083038}
}

@article{Anthony-Petersen:2025,
    author = {Anthony-Petersen, R. and others},
    title = {{Low energy backgrounds and excess noise in a two-channel low-threshold calorimeter}},
    journal = {Appl. Phys. Lett.},
    volume = {126},
    number = {10},
    pages = {102601},
    year = {2025},
    month = {03},
    url = {https://doi.org/10.1063/5.0247343},
}

@article{Antman:1966,
title = {{Measurements of the Fano factor and the energy per hole-electron pair in germanium}},
author = {Antman, S O W and Landis, D A and Pehl, R H},
journal = {Nucl. Instrum. Meth.},
volume = {40},
number = {2},
pages = {272-276},
year = {1966},
issn = {0029-554X},
url = {https://doi.org/10.1016/0029-554X(66)90386-7}
}

@article{Aralis:2020,
  title = {{Constraints on dark photons and axionlike particles from the SuperCDMS Soudan experiment}},
  author = {Aralis, T. and others},
  collaboration = {SuperCDMS Collaboration},
  journal = {Phys. Rev. D},
  volume = {101},
  issue = {5},
  pages = {052008},
  numpages = {14},
  year = {2020},
  month = {Mar},
  publisher = {American Physical Society},
  url = {https://doi.org/10.1103/PhysRevD.101.052008}
}

@article{Aralis:2021,
  title = {{Erratum: Constraints on dark photons and axionlike particles from the SuperCDMS Soudan experiment [Phys. Rev. D 101, 052008 (2020)]}},
  author = {Aralis, T and others},
  collaboration = {SuperCDMS Collaboration},
  journal = {Phys. Rev. D},
  volume = {103},
  issue = {3},
  pages = {039901},
  numpages = {3},
  year = {2021},
  month = {Feb},
  publisher = {American Physical Society},
  url = {https://doi.org/10.1103/PhysRevD.103.039901}
}

@article{Armatol:2025,
  title = {{Characterization of mini-CryoCube detectors from the RICOCHET experiment commissioning at the Institut Laue-Langevin}},
  author = {Armatol, A. and others},
  collaboration = {RICOCHET Collaboration},
  journal = {Phys. Rev. D},
  volume = {112},
  issue = {11},
  pages = {112019},
  numpages = {16},
  year = {2025},
  month = {Dec},
  publisher = {American Physical Society},
  url = {https://link.aps.org/doi/10.1103/7xy6-jq3c}
}

@article{Armengaud:2017,
url = {https://doi.org/10.1088/1748-0221/12/08/P08010},
year = {2017},
month = {aug},
volume = {12},
number = {08},
pages = {P08010},
author = {Armengaud, E. and others},
collaboration = {EDELWEISS Collaboration},
title = {{Performance of the EDELWEISS-III experiment for direct dark matter searches}},
journal = {JINST},
}

@article{Arnaud:2020,
  title = {{First Germanium-Based Constraints on Sub-MeV Dark Matter with the EDELWEISS Experiment}},
  author = {Arnaud, Q. and others},
  collaboration = {EDELWEISS Collaboration},
  journal = {Phys. Rev. Lett.},
  volume = {125},
  issue = {14},
  pages = {141301},
  numpages = {6},
  year = {2020},
  month = {Oct},
  publisher = {American Physical Society},
  url = {https://doi.org/10.1103/PhysRevLett.125.141301}
}

@article{Augier:2024,
  author       = {Augier, C and others},
  collaboration = {RICOCHET Collaboration},
  title        = {{First demonstration of 30\,eVee ionization energy resolution with Ricochet germanium cryogenic bolometers}},
  journal      = {Eur. Phys. J. C},
  volume       = {84},
  number       = {2},
  pages        = {186},
  year         = {2024},
  month        = feb,
  url          = {https://doi.org/10.1140/epjc/s10052-024-12433-1}
}

@article{Baxter:2025,
   author = {Baxter, Daniel and others},
   title = {{Low-Energy Backgrounds in Solid-State Phonon and Charge Detectors}}, 
   journal= {Annual Review of Nuclear and Particle Science},
   year = {2025},
   volume = {75},
   number = {Volume 75, 2025},
   pages = {301-326},
   url = {https://www.annualreviews.org/content/journals/10.1146/annurev-nucl-121423-100849},
   publisher = {Annual Reviews},
   issn = {1545-4134},
   type = {Journal Article},
  }

@article{Bilger:1967,
  title = {{Fano Factor in Germanium at 77\textdegree K}},
  author = {Bilger, Hans R.},
  journal = {Phys. Rev.},
  volume = {163},
  issue = {2},
  pages = {238--253},
  numpages = {0},
  year = {1967},
  month = {Nov},
  publisher = {American Physical Society},
  url = {https://doi.org/10.1103/PhysRev.163.238}
}

@article{Bloch:2025,
  title = {{SENSEI at SNOLAB: Single-Electron Event Rate and Implications for Dark Matter}},
  author = {Bloch, Itay M. and others},
  collaboration = {SENSEI Collaboration},
  journal = {Phys. Rev. Lett.},
  volume = {134},
  issue = {16},
  pages = {161002},
  numpages = {7},
  year = {2025},
  month = {Apr},
  publisher = {American Physical Society},
  url = {https://doi.org/10.1103/PhysRevLett.134.161002}
}

@article{Broniatowski:2006,
title = {{Dead layer and degradation effects in cryogenic germanium detectors for dark matter search}},
journal = {Nucl. Instrum. Meth. A},
volume = {559},
number = {2},
pages = {402-404},
year = {2006},
url = {https://doi.org/10.1016/j.nima.2005.12.172},
author = {Broniatowski, A and Censier, B and Juillard, A and Berg\'e, L}
}

@article{Burlac:2025,
title = {{Early results of the LEGEND-200 experiment}},
journal = {Nucl. Instrum. Meth. A},
volume = {1080},
pages = {170779},
year = {2025},
url = {https://doi.org/10.1016/j.nima.2025.170779},
author = {Nina Burlac}
}

@misc{COMSOL,
  title = {{COMSOL -- Software for Multiphysics Simulation}},
  author = {},
  howpublished = {\url{https://www.comsol.com/}},
  note = {Accessed: 2025-09-25}}

@article{Crauste:2013,
  title     = {{Effect of annealing on the superconducting properties of a-NbSi thin films}},
  author    = {Crauste, O. and others},
  journal   = {Phys. Rev. B},
  volume    = {87},
  number    = {14},
  pages     = {144514},
  numpages  = {11},
  year      = {2013},
  publisher = {American Physical Society},
  url       = {https://doi.org/10.1103/PhysRevB.87.144514}
}

@article{Cusini:2022,
AUTHOR={Cusini, Iris  and others},     
TITLE={{Historical Perspectives, State of art and Research Trends of Single Photon Avalanche Diodes and Their Applications (Part 1: Single Pixels)}},
JOURNAL={Front. Phys.},
VOLUME={10},
YEAR={2022},
url={https://doi.org/10.3389/fphy.2022.906675},
}

@article{Fano:1947,
  author    = {Fano, U.},
  title     = {{Ionization Yield of Radiations. II. The Fluctuations of the Number of Ions}},
  journal   = {Phys. Rev.},
  volume    = {72},
  pages     = {26--29},
  year      = {1947},
  publisher = {American Physical Society},
  url       = {https://doi.org/10.1103/PhysRev.72.26}
}

@article{Geng:2024,
url = {https://doi.org/10.1088/1475-7516/2024/07/009},
year = {2024},
month = {jul},
publisher = {IOP Publishing},
volume = {2024},
number = {07},
pages = {009},
author = {Geng, X.P. and others},
collaboration = {CDEX Collaboration},
title = {{Projected WIMP sensitivity of the CDEX-50 dark matter experiment}},
journal = {JCAP},
}

@article{Genz:1971,
  title = {{Multiwire-Proportional-Counter Measurement of the $\frac{M}{L}$ Orbital-Electron-Capture Ratio in $^{71}\mathrm{Ge}$ Decay}},
  author = {Genz, H. and Renier, J. P. and Pengra, J. G. and Fink, R. W.},
  journal = {Phys. Rev. C},
  volume = {3},
  issue = {1},
  pages = {172--179},
  numpages = {0},
  year = {1971},
  month = {Jan},
  publisher = {American Physical Society},
  url = {https://doi.org/10.1103/PhysRevC.3.172}
}

@article{Ichimura:2023,
    author = {Ichimura, K and others},
    title = {{Development of a low-background HPGe detector at Kamioka Observatory}},
    journal = {Prog. Theor. Exp. Phys.},
    volume = {2023},
    number = {12},
    pages = {123H01},
    year = {2023},
    month = {11},
    url = {https://doi.org/10.1093/ptep/ptad136},
}

@article{Irwin:1995,
    author = {Irwin, K. D.},
    title = {{An application of electrothermal feedback for high resolution cryogenic particle detection}},
    journal = {Appl. Phys. Lett.},
    volume = {66},
    number = {15},
    pages = {1998-2000},
    year = {1995},
    month = {04},
    url = {https://doi.org/10.1063/1.113674},

}

@Inbook{Irwin:2005,
author={Irwin, K.D. and Hilton, G.C.},
editor={Enss, Christian},
title={{Transition-Edge Sensors}},
bookTitle={Cryogenic Particle Detection},
year={2005},
publisher={Springer Berlin Heidelberg},
address={Berlin, Heidelberg},
pages={63--150},
url={https://doi.org/10.1007/10933596_3}
}

@article{Kennard:2026,
   title={{Performance of a SuperCDMS HVeV detector with Sub-eV energy resolution and single charge-sensitivity}},
journal = {Nucl. Instrum. Meth. A},
volume = {1091},
pages = {171753},
year = {2026},
issn = {0168-9002},
url = {https://www.sciencedirect.com/science/article/pii/S0168900226004791},
author = {Kyle Kennard and others},
}

@article{Lewin:1996,
  author       = {J. D. Lewin and P. F. Smith},
  title        = {{Review of mathematics, numerical factors, and corrections for dark matter experiments based on elastic nuclear recoil}},
  journal      = {Astroparticle Physics},
  volume       = {6},
  number       = {1},
  pages        = {87--112},
  year         = {1996},
  month        = dec,
  issn         = {0927-6505},
  publisher    = {Elsevier}
}

@article{Li:2025,
  title        = {{New physics versus quenching factors in Coherent Neutrino Scattering}},
  author       = {Li, Yulun and Herrera, Gonzalo and Huber, Patrick},
  journal      = {JHEP},
  volume       = {2025},
  number       = {11},
  pages        = {22},
  year         = {2025},
  url          = {https://doi.org/10.1007/JHEP11(2025)022},
}

@article{Lindhard:1963,
  author  = {J. Lindhard and V. Nielsen and M. Scharff and P. V. Thomsen},
  title   = {{Integral Equations Governing Radiation Effects (Notes on Atomic Collisions III)}},
  journal = {Mat. Fys. Medd. Dan. Vid. Selsk.},
  volume  = {33},
  number  = {10},
  pages   = {1--42},
  year    = {1963}
}

@article{Liu:2024,
title = {{Frontiers and challenges in silicon-based single-photon avalanche diodes and key readout circuits}},
journal = {Microelectron. J.},
volume = {147},
pages = {106165},
year = {2024},
url = {https://doi.org/10.1016/j.mejo.2024.106165},
author = {Yang Liu and others},
}

@article{LHote:2000,
author = {L’Hôte, D. and Navick, X. F. and Tourbot, R. and Mangin, J. and Pesty, F.},
title = {{Charge and heat collection in a 70 g heat/ionization cryogenic detector for dark matter search}},
journal = {J. Appl. Phys.},
volume = {87},
number = {3},
pages = {1507-1521},
year = {2000},
month = {02},
url = {https://doi.org/10.1063/1.372042},

}

@article{Looker:2015,
title = {{Leakage current in high-purity germanium detectors with amorphous semiconductor contacts}},
journal = {Nucl. Instrum. Meth. A},
volume = {777},
pages = {138-147},
year = {2015},
issn = {0168-9002},
url = {https://doi.org/10.1016/j.nima.2014.12.104},
author = {Q. Looker and M. Amman and K. Vetter},
}

@article{Lowe:1997,
title = {{Measurements of Fano factors in silicon and germanium in the low-energy X-ray region}},
journal = {Nucl. Instrum. Meth. A},
volume = {399},
number = {2},
pages = {354-364},
year = {1997},
issn = {0168-9002},
url = {https://doi.org/10.1016/S0168-9002(97)00965-0},
author = {B.G. Lowe},
}

@article{Luke:1988,
    author = {Luke, P. N.},
    title = {{Voltage‐assisted calorimetric ionization detector}},
    journal = {J. Appl. Phys.},
    volume = {64},
    number = {12},
    pages = {6858-6860},
    year = {1988},
    month = {12},
    url = {https://doi.org/10.1063/1.341976},
}

@article{Macfarlane:1974,
  title = {{Fine Structure in the Absorption-Edge Spectrum of Ge}},
  author = {Macfarlane, G. G. and McLean, T. P. and Quarrington, J. E. and Roberts, V.},
  journal = {Phys. Rev.},
  volume = {108},
  issue = {6},
  pages = {1377--1383},
  numpages = {0},
  year = {1957},
  month = {Dec},
  publisher = {American Physical Society},
  url = {https://doi.org/10.1103/PhysRev.108.1377}
}

@article{Marnieros:2012,
  author       = {Marnieros, S. and others},
  title        = {{Electron-Phonon Decoupling NbSi CMB Bolometers}},
  journal      = {J. Low Temp. Phys.},
  volume       = {167},
  number       = {5-6},
  pages        = {846-851},
  year         = {2012},
  url          = {https://doi.org/10.1007/s10909-012-0532-8}
}

@article{Mei:2024,
  author       = {Dongming Mei},
  title        = {{Exploring the Potential of Residual Impurities in Germanium Detectors for MeV-Scale Dark Matter Detection}},
  journal      = {J. Low Temp. Phys.},
  volume       = {216},
  pages        = {522--537},
  year         = {2024},
  url          = {https://doi.org/10.1007/s10909-024-03059-4}
}

@article{Miller:1955,
  author       = {S. L. Miller},
  title        = {{Avalanche Breakdown in Germanium}},
  journal      = {Phys. Rev.},
  year         = {1955},
  volume       = {99},
  pages        = {1234--1241},
  url          = {https://doi.org/10.1103/PhysRev.99.1234}
}

@article{Na:2024,
  author       = {Na, Neil and others},
  title        = {{Room temperature operation of germanium–silicon single-photon avalanche diode}},
  journal      = {Nature},
  year         = {2024},
  volume       = {627},
  pages        = {295--300},
  url          = {https://doi.org/10.1038/s41586-024-07076-x},
}

@article{Neganov:1985,
  author={Neganov, B.S. and Trofimov, V.N.},
  title={{{USSR} patent No 1037771}},
  journal={Otkrytia i Izobreteniya},
  volume={146},
  pages={215},
  year={1985}}

@article{Norman:2024,
  title = {{Half-life of $^{71}\mathrm{Ge}$ and the gallium anomaly}},
  author = {Norman, E. B. and others},
  journal = {Phys. Rev. C},
  volume = {109},
  issue = {5},
  pages = {055501},
  numpages = {5},
  year = {2024},
  month = {May},
  publisher = {American Physical Society},
  url = {https://doi.org/10.1103/PhysRevC.109.055501}
}

@article{Olivieri:2009,
    author = {Olivieri, E. and Broniatowski, A. and Domange, J. and Defay, X. and Chapellier, M. and Dumoulin, L.},
    title = {{Space‐and‐surface charge neutralization of cryogenic Ge detectors using infrared LEDs}},
    journal = {AIP Conf. Proc.},
    volume = {1185},
    number = {1},
    pages = {310-313},
    year = {2009},
    month = {12},
    url = {https://doi.org/10.1063/1.3292566},
}

@article{Olivieri:2017,
title = {{Vibrations on pulse tube based Dry Dilution Refrigerators for low noise measurements}},
journal = {Nucl. Instrum. Meth. A},
volume = {858},
pages = {73-79},
year = {2017},
issn = {0168-9002},
url = {https://doi.org/10.1016/j.nima.2017.03.045},
author = {E. Olivieri and J. Billard and M. {De Jesus} and A. Juillard and A. Leder},
}

@article{Ponce:2020,
  title = {{Measuring the impact ionization and charge trapping probabilities in SuperCDMS HVeV phonon sensing detectors}},
  author = {Ponce, F. and others},
  journal = {Phys. Rev. D},
  volume = {101},
  issue = {3},
  pages = {031101(R)},
  numpages = {5},
  year = {2020},
  month = {Feb},
  publisher = {American Physical Society},
  url = {https://link.aps.org/doi/10.1103/PhysRevD.101.031101}
}

@article{Romani:2018,
    author = {Romani, R. K. and others},
    title = {{Thermal detection of single e-h pairs in a biased silicon crystal detector}},
    journal = {Appl. Phys. Lett.},
    volume = {112},
    number = {4},
    pages = {043501},
    year = {2018},
    month = {01},
    url = {https://doi.org/10.1063/1.5010699},
}

@article{Shutt:1992,
  title = {{Simultaneous high resolution meausurement of phonons and ionization created by particle interactions in a 60 g germanium crystal at 25 mK}},
  author = {Shutt, T. and others},
  journal = {Phys. Rev. Lett.},
  volume = {69},
  issue = {24},
  pages = {3531--3534},
  numpages = {0},
  year = {1992},
  month = {Dec},
  publisher = {American Physical Society},
  url = {https://doi.org/10.1103/PhysRevLett.69.3531}
}

@article{Tiffenberg:2017,
  title = {{Single-Electron and Single-Photon Sensitivity with a Silicon Skipper CCD}},
  author = {Tiffenberg, Javier and others},
  journal = {Phys. Rev. Lett.},
  volume = {119},
  issue = {13},
  pages = {131802},
  numpages = {6},
  year = {2017},
  month = {Sep},
  publisher = {American Physical Society},
  url = {https://doi.org/10.1103/PhysRevLett.119.131802}
}

@article{Wei:2017,
url = {https://doi.org/10.1088/1748-0221/12/04/P04022},
year = {2017},
month = {apr},
volume = {12},
number = {04},
pages = {P04022},
author = {Wei, W.-Z. and Mei, D.-M.},
title = {{Average energy expended per e-h pair for germanium-based dark matter experiments}},
journal = {JINST},
}

@article{Wei:2020,
  author       = {Wei, W.-Z. and Panth, R. and Liu, J. and Mei, H. and Mei, D.-M. and Wang, G.-J.},
  title        = {{The impact of the charge barrier height on Germanium (Ge) detectors with amorphous-Ge contacts for light dark matter searches}},
  journal      = {Eur. Phys. J. C},
  volume       = {80},
  number       = {472},
  year         = {2020},
  url          = {https://doi.org/10.1140/epjc/s10052-020-8029-0}
}

@article{Wilson:2024,
  title = {{Improved modeling of detector response effects in phonon-based crystal detectors used for dark matter searches}},
  author = {Wilson, M. J. and others},
  journal = {Phys. Rev. D},
  volume = {109},
  issue = {11},
  pages = {112018},
  numpages = {20},
  year = {2024},
  month = {Jun},
  publisher = {American Physical Society},
  url = {https://link.aps.org/doi/10.1103/PhysRevD.109.112018}
}

@article{Zurauskas:1992,
author = {Žurauskas, S. and Dargys, A. and Žurauskienė, N.},
title = {{Field Ionization of Shallow Donors in Germanium}},
journal = {Physica Status Solidi (B)},
volume = {173},
number = {2},
pages = {647-660},
url = {https://doi.org/10.1002/pssb.2221730217},
year = {1992}
}

\end{document}